# Toward ultra-efficient high-fidelity prediction of bed morphodynamics of large-scale meandering rivers using a novel LES-trained machine learning approach

Zexia Zhang[1], Mehrshad Gholami Anjiraki[1], Hossein Seyedzadeh[1], Fotis Sotiropoulos[2], and Ali Khosronejad[1*]

[1]Civil Engineering Department, Stony Brook University, Stony Brook, NY 11794, USA
[2]Mechanical and Nuclear Engineering, Virginia Commonwealth University, Richmond, VA 23284, USA

[*]Corresponding author, Email: ali.khosronejad@stonybrook.edu

**Abstract**
Flood-induced deformation of the bed topography of fluvial meandering rivers could lead to river bank displacement, structural failure of the infrastructures, and the propagation of scour or deposition features. The assessment of sediment transport in large-scale meanders is, therefore, a key environmental issue. High-fidelity numerical models provide powerful tools for such assessments. However, high-fidelity simulations of large-scale rivers using the coupled flow and morphodynamics modules can be computationally expensive, owing to the costly two-way coupling between turbulence and bed morphodynamics. This study seeks to present a novel machine learning approach, which is trained using coupled large-eddy simulation (LES) and morphodynamics results. The proposed machine learning approach predicts bed shear stress and equilibrium bed morphology of large-scale meanders under bankfull flow conditions at a fraction of the cost of coupled LES-morphodynamics. We developed and evaluated the performance of a convolutional neural network autoencoder (CNNAE) algorithm to generate high-fidelity bed shear stress and equilibrium morphology of large-scale meandering rivers. The CNNAE algorithm utilizes instantaneous shear stress distribution and change of bed elevation of high-fidelity simulation results, along with geometric parameters of meanders as inputs to predict mean bed shear stress distribution and equilibrium bed elevation of rivers. The results demonstrated the feasibility, accuracy, and efficiency of the proposed CNNAE algorithm.

## 1. Introduction
Predicting the interaction between turbulent river flow and bed topography in the field-scale meandering rivers is essential for various river engineering and geoscience problems. For example, such predictions allow researchers and practicing engineers to gain insight into the evolution of the riverbeds and, thus, to design effective flood management and mitigation strategies. Also, sediment transport affects the deformation and migration of river channels, which is important for navigation and maintaining infrastructure such as bridges and dams. A commonly utilized engineering tool for such predictions at large-scale rivers is the high-fidelity numerical models that are based on computational fluid dynamics (CFD) simulations (Behzad et al. 2023; Bigdeli et al. 2023; Bourgoin et al. 2021; Flora et al. 2021; Flora and Khosronejad 2021, 2022; Khosronejad et al. 2019b, 2020b; c, a; e; Khosronejad and Sotiropoulos 2017, 2020; Nian et al. 2021b; a;

Yazdanfar et al. 2021). Compared to the high-fidelity models, the emerging artificial intelligence (AI) based machine learning algorithms seem to provide a more efficient approach for the prediction of bed deformation in large-scale meandering rivers. This study attempts to develop, train, and validate the machine learning algorithms that can generate high-fidelity bed topography of large-scale meandering rivers under bankfull conditions and at a fraction of the cost associated with the high-fidelity CFD models.

Jain (2001) was among the first to employ artificial intelligence (AI) methods to predict river sediment dynamics. He adopted a feedforward neural network to connect the dynamics of river stage, discharge, and sediment concentration in the Mississippi River. Building on this work, Rai and Mathur (2008) developed artificial neural networks (ANN) models to predict event-based and time-dependent changes in sediment yield by integrating sediment flux, runoff, and rainfall data. Jothiprakash and Garg (2009) focused on predicting sediment retention within reservoirs using a Multi-Layer Perceptron (MLP) and the backpropagation technique, drawing on annual rainfall, inflow, and waterway capacity. Taking an alternative route, Cigizoglu (2004) explored the potential of MLP for predicting suspended sediment discharge, skillfully leveraging extensive daily flow and sediment data collected from the Schuylkill River. The scope of exploration was further broadened by Melesse et al. (2011), who conducted a comprehensive evaluation of AI techniques encompassing both linear and non-linear regressions, aiming to forecast suspended sediment loads across various river systems by incorporating factors like rainfall, discharge, and sediment load data. Concurrently, Khosravi et al., (2020) embarked on the task of predicting bedload sediment transport rates using data mining techniques and laboratory flume data, thereby enriching our understanding of sediment dynamics.

Furthermore, Zhang et al. (2022) explored adaptive neuro-fuzzy inference system (ANFIS) for predicting instantaneous three-dimensional (3D) velocity fields of flood flow in large-scale rivers and validated their results against those of large-eddy simulation (LES). They also introduced a recurrent method combining LES and ANFIS to enhance multi-time-step predictions. Utilizing ANN-based models, Bhattacharya et al. (2012) predicted suspended sediment concentration profiles by incorporating data on wind, wave, and bed shear stress.

In addition, a multitude of cutting-edge methodologies are explored through capabilities of convolutional neural networks (CNN). Kabir et al. (2020) predicted water depths in Carlisle, UK, through the application of a CNN model, trained by a two-dimensional (2D) hydraulic model. They further enriched their CNN predictions via a comparative analysis against the support vector regression method. Continuing along this trajectory, Collins et al. (2020) adopted a 2D CNNs to predict nearshore bathymetry, utilizing synthesized imagery to capture the wave dynamics. Expanding the horizon even further, Takechi et al. (2021) employed a CNN-based image recognition approach to categorize riverbed particle sizes through aerial photography, paralleled by the groundbreaking work of Ghanbari and Antoniades (2022), who introduced a pioneering one-dimensional (1D) CNN architecture for the prediction of sediment particle sizes. Their model was duly validated through the rigorous examination of hyperspectral imagery. Similarly, Lang et al. (2021) harnessed the potential of CNNs to study the distribution of grain sizes within gravel bars, utilizing drone-captured images as a rich source of input data. In another endeavor, Zhang et al. (2022a) developed encoder-decoder CNNs with the ability to produce 3D depictions of turbulent flood flow in large-scale rivers with wall-mounted bridge piers. These models were designed to forecast the time-averaged flow patterns by leveraging instantaneous and time-averaged LES data, encompassing variations with and without a physical constraint that guaranteed a divergence-free conditions within the time-averaged flow field. Moreover, de Melo et al. (2022) simulated bathymetric changes through CNN, incorporating input variables derived from hydro- and morpho-dynamic numerical models to predict erosion and sedimentation variations. Collectively, these innovative endeavors illuminate the ever-evolving landscape of AI-driven predictions within the realm of sediment transport and bed deformation.

In a complementary manner, Genç et al. (2015) combined ANN with other machine learning methods to forecast shear stress distribution in small streams, factoring in parameters like water surface slope and flow velocity. Meanwhile, Nagy et al. (2002) leveraged a multilayer feedforward neural network along with a backpropagation training algorithm to predict sediment load, utilizing input variables like tractive shear stress and Froude number. This compilation of studies collectively demonstrates the power of these techniques in enhancing our understanding of shear dynamics and sediment transport within hydraulic systems. Furthermore, a diverse array of innovative approaches enriches the landscape of research in this domain. Roushangar et al. (2014) employed AI-driven models that incorporated Gene Expression Programming (GEP) and ANFIS method to estimate bed material load based on sediment data, complementing the insights gained from stream power and shear stress-based empirical models. Likewise, shear stress distribution within a rectangular channel was delved into by Lashkar-Ara et al. (2021), using Tsallis entropy, Genetic Programming (GP), and ANFIS techniques to explore parameters related to channel walls and bed. Harasti et al. (2023) employed a non-linear regression model to predict scour depth around bridge piers, incorporating variables like effective pier width and flow depth alongside local and critical bed shear stress. Similarly, Kitsikoudis et al. (2014) formulated a series of sediment transport equations by a fusion of machine learning methods, incorporating data on shear stress from field observations and laboratory experiments. Shakya et al. (2023) tackled predicting total sediment load in alluvial channels, highlighting the paramount role of dynamic properties like channel discharge, friction slope, and bed shear stress. Meanwhile, Mohanta et al. (2021) introduced AI methodologies to calculate shear force proportion in two-stage meandering channels, a vital factor for floodplain conveyance. Sheikh Khozani et al. (2020) explored shear stress distribution in compound channels with varying floodplain widths and flow depths, leveraging AI methods such as multivariate adaptive regression spline (MARS), group method of data handling neural network (GMDH-NN), and gene-expression programming (GEP).

The CFD based, high-fidelity simulations of the bed evolution of large-scale meandering rivers during a flood or bank full flow conditions have proven to be technically challenging and computationally expensive tasks (Khosronejad et al. 2023). It is technically challenging owing to the complexities involved in the modeling of meandering river planform geometries and the intricate interaction between turbulence and morphodynamics. The couple flow and sediment transport simulations of real-life meandering rivers are quite expensive because of the sheer number of computational grid nodes required to resolve the large domain of the meandering rivers could easily reach several kilometers in length and hundreds of meters in width. Even though the wall modeling can alleviate some of the high computational cost, the number of computational grid nodes required for such high-fidelity simulations would easily exceed hundreds of millions. The coupling between the flow and morphodynamics could yet, at best, double the computational cost of the simulations. We note that the couple simulations are required to be continued for an extended period of time until the bed morphology of the rivers is at dynamics equilibrium when the major bed changes are dynamic but stable. Such a coupled hydro- and morpho-dynamics simulation, on average, would take about 200,000 CPU hours(Khosronejad et al. 2023). This study seeks to develop and validate an efficient approach to reduce the computational cost of such high-fidelity modeling while maintaining the accuracy of the predictions by enforcing adequate physical constraints.

Herein, we proposed a novel convolutional neural network autoencoder (CNNAE) algorithm to predict the time-averaged shear stress distribution and equilibrium bed elevation of mobile riverbed in large scale meandering rivers. We performed hydrodynamics and morphodynamics simulations in sixteen virtual meandering rivers with different geometries to generate training and validation data for the CNNAE algorithm. As input, the developed CNNAE algorithm employs the instantaneous snapshots of shear stress distribution and change of bed elevation at the very beginning of the coupled hydrodynamics-morphodynamics simulation, i.e., the couple simulation results of the first time step, along with the geometry parameters, e.g., streamwise and spanwise curvilinear coordinates, and local curvature. As output, the CNNAE algorithm reconstructs the time-averaged shear stress distribution and the bed elevation of the river at its dynamic equilibrium. The developed CNNAE algorithm is trained using the coupled LES-morphodynamics simulation results of five large-scale meandering rivers. The trained CNNAE algorithm is then validated against the coupled LES-morphodynamics simulation results of 11 large-scale meandering rivers. The rivers used during the training and validation are distinctly different in terms of their planform geometry, scale and hydraulic characteristics. The comparison of the CNNAE predictions against the coupled LES-morphodynamics simulation results marked a accuracy and efficiency of the proposed CNNAE algorithm, which costs less than two percent of the coupled LES-morphodynamics models.

This paper is organized as follows. Section 2 presents the numerical methods for coupled hydrodynamics and morphodynamics simulations. Section 3 describes the computational details of the numerical simulations of the sixteen meandering rivers to produce training and validation data for the machine learning algorithm. Section 4 introduces the proposed CNNAE algorithm, and the workflow of the numerical processes of the training and validation phases. Lastly, in Section 5, we present and discuss the prediction results of the machine learning algorithm and compare them with the high-fidelity coupled LES-morphodynamics results. Finally, the findings of this study are concluded in Section 6.

## 2. Numerical Methods

Herein, we provide a concise overview of the equations that govern the hydrodynamics and morphodynamics of fluvial river systems within our simulation modeling framework. For a comprehensive elaboration on the mathematical foundation of our models, as well as the specific technique used to couple the hydrodynamics and morphodynamics modules, we refer the reader to (Khosronejad et al. 2011, 2012, 2013, 2014, 2019a; c, 2020d; Khosronejad and Sotiropoulos 2014; Yang et al. 2017).

### 2.1. The hydrodynamic model

The bankfull river flow field is obtained using the spatially averaged continuity and Navier-Stokes equations, which resolved the 3D instantaneous incompressible turbulent flow. These equations written in compact tensor notation and curvilinear coordinates read as follows ($i, j, k, l = 1,2,3$):

$$J \frac{\partial U_j}{\partial \xi_j} = 0 \tag{1}$$

$$\frac{1}{J}\frac{\partial U_i}{\partial t} = \frac{\xi_l^i}{J}\left[-\frac{\partial(U_j u_l)}{\partial \xi_j} + \frac{\partial}{\partial \xi_j}\left(\nu \frac{G^{jk}}{J}\frac{\partial u_l}{\partial \xi_k}\right) - \frac{1}{\rho}\frac{\partial}{\partial \xi_j}\left(\frac{\xi_l^i p}{J}\right) - \frac{1}{\rho}\frac{\partial \tau_{lj}}{\partial \xi_j}\right] \tag{2}$$

where $\xi^i$ are the curvilinear coordinates, $J$ is the Jacobian of the geometric transformation, $u_i$ is the $i^{th}$ component of the velocity vector in Cartesian coordinates, $U^i = (\xi^i_m/J)u_m$ is the contravariant volume flux, $v$ is the kinematic viscosity of water, $G^{jk} = \xi^i_l \xi^k_l$ are the components of the contravariant metric tensor, $p$ is the pressure, $\rho$ is water density, and $\tau_{lj}$ is the subgrid stress tensor of the LES. The velocity field in the filtered Navier-stocks equation is decomposed to resolved and unresolved components, and the subgrid stress tensor in the momentum equations is used to represent the unresolved stress terms, which are modeled by the dynamic Smagorinsky sub-grid scale (SGS) model in the following manner:

$$\tau_{ij} - \frac{\delta_{ij}}{3}\tau_{kk} = -2\mu_t \overline{S_{ij}} \qquad (3)$$

where the overbar indicates the grid filtering operation, $\overline{S_{ij}}$ is the filtered strain-rate tensor, $\mu_t = C_s \Delta^2 |\bar{S}|$, is the eddy viscosity, $C_s$ is the Smagorinsky constant, also $|\bar{S}| = \sqrt{2\overline{S_{ij}S_{ij}}}$, $\Delta = J^{-1/3}$ is the filter size obtained from the box filter, and $\delta_{ij}$ is the Kronecker delta function. Employing the dynamic Smagorinsky Subgrid-Scale (SGS) model, the constant $C_s$ varies over time and space, influenced by the flow field. This approach is particularly well-suited for turbulent flows with high Reynolds numbers, commonly observed in natural river channels. (See Kang et al., 2011).

The terms related to convection, divergence, pressure gradient, and viscous in the governing equations underwent spatial discretization on a staggered/non-staggered computational grid system, using a central numerical scheme of second-order accuracy. Time derivatives were handled with second-order backward differencing, and the integration in time followed a fractional step approach that was second-order accurate. To address the momentum equations, a Jacobian-free Newton-Krylov solver was employed in conjunction with the fractional step method. Additionally, in solving the Poisson equation, we utilized a solver based on the generalized minimal residual method, which was further enhanced by employing a multigrid as a preconditioner (for more details, see Kang et al., (2011)).

The LES was implemented in the framework of the Curvilinear Immersed Boundary (CURVIB) method, enabling us to model highly complicate geometric configurations, such as the meandering rivers with deformable beds. It's important to emphasize that within the CURVIB method the background computational domain for each meandering river closely conforms to the river's curvature. This domain is discretized using a grid system that mirrors the river's curvature. The riverbanks and the interface between sediment and water, integrated into the background grid system, are represented using unstructured triangular grids. The hydrodynamic governing equations were computed at the nodes of the background grid within the fluid phase. Boundary conditions were applied at fluid nodes located in the immediate vicinity of both the sediment/water interface and riverbanks. To elucidate further, the nodes along the boundary are termed as the immersed boundary (IB) nodes. Additionally, the computational nodes within the unstructured triangular grid system — specifically encompassing the riverbanks and sediment layer — were omitted from the computations (Khosronejad et al. 2019b, 2020b; c).

## 2.2. Morphodynamics
This section outlines the governing equations of bed deformation and suspended sediment transportation. The non-equilibrium Exner-Polya equation that describes the sediment mass balance governs the time variation of the river bed elevation as follows (Khosronejad et al. 2020e; Khosronejad and Sotiropoulos 2017):

$$(1 - \varphi)\frac{\partial z_b}{\partial t} = -\nabla \cdot \boldsymbol{q}_{\text{BL}} + D_b - E_b \tag{4}$$

where $z_b$ is the bed elevation, $\boldsymbol{q}_{\text{BL}}$ is the bed-load flux vector, $\varphi$ is the sediment martial porosity (= 0.4), $\nabla$ is the divergence operator, $D_b$ represents the rate of net sediment deposited from the suspension onto the bed, while $E_b$ stands for the rate of net sediment which is picked up from the bed and enters the flow domain, also referred to as the particle pick-up rate. The flux vector for bed load within the bed load layer is determined as follows:

$$\boldsymbol{q}_{\text{BL}} = C_{\text{BL}} \delta_{\text{BL}} \boldsymbol{u}_{\text{BL}} \tag{5}$$

where $C_{\text{BL}}$ refers to the sediment concentration within the bed load layer, which has a thickness of $\delta_{\text{BL}}$. $\boldsymbol{u}_{\text{BL}}$ represents the velocity vector parallel the bed surface at the interface between water and sediment (i.e., at the top of the bed load layer). The values for bed-load sediment concentration ($C_{\text{BL}}$) and bed-load layer thickness ($\delta_{\text{BL}}$) were computed using van Rijn's (1993) equations, which depend on the local bed shear stress and the threshold at which sediment particles start moving. The critical bed shear stress is determined by applying Shield's criterion (Shields 1936), utilizing the parameterized form of Shield's curve as described by van Rijn (1993). This calculation initially considers the critical shear stress for a flat bed and is subsequently adjusted to account for beds with both transverse and longitudinal slopes.

The calculated values for the net rates of sediment deposition ($D_b$) and entrainment ($E_b$) over the mobile bed are determined as follows:

$$D_b = w_s C_b \tag{6}$$
$$E_b = w_s C_{\text{BL}} \tag{7}$$

where $w_s$ represents the settling speed of the non-spherical sediment particles, determined using van Rijn's formula (van Rijn, 1993). $C_b$ stands for the sediment concentration right above the bed load layer. The deposition rate is connected to the sediment material that moves vertically from the flow area onto the mobile bed. To calculate $C_b$ from the suspended sediment concentration field C within the flow area, we applied a quadratic interpolation method.

In the case of a dilute sediment-water mixture, where the volumetric sediment concentration is below $O$ (0.01), we model the suspended sediment concentration field within the flow domain by employing the subsequent convection-diffusion equation:

$$\frac{1}{J}\frac{\partial(\rho C)}{\partial t} + \frac{\partial\left(\rho C(U_j - W_j \delta_{ij})\right)}{\partial \xi_j} = \frac{\partial}{\partial \xi_j}\left(\left(\frac{v}{S_L} + \frac{v_t}{S_T}\right)\frac{G^{jk}}{J}\frac{\partial C}{\partial \xi_k}\right) \tag{8}$$

where, $W_j = (\xi_3^j/J)w_s$ represents the contravariant volume flux of the suspended sediment concentration due to particle settling in the flow area. $S_L$ (= 700), stands for the laminar Schmidt number, $S_T$ (= 0.75) denotes the turbulent Schmidt number and $v_t$ signifies the kinematic eddy viscosity. A second-order central differencing numerical method is obtained to discretize the convection-diffusion equation and the equation is solved using the fully implicit Jacobian free Newton approach.

Importantly, in each time-step of the numerical algorithm, we check the computed bed surface slopes at the end of each hydro-morphodynamics step in order to prevent unrealistic slopes at the interface of fluid and sediment. We ensure that the calculated surface slope does not surpass the angle of repose for the sediment material. To achieve this, we implement a mass-conserving sand-slide module. This module identifies any unrealistic local slopes and redistributes sediment mass among adjacent bed cells, ensuring that no local bed surface slope exceeds the angle of repose (see, e.g., (Khosronejad et al. 2011; Khosronejad and Sotiropoulos 2014, 2017).

## 2.3. Coupling of Hydrodynamics and Morphodynamics

To model the coupled interactions between hydrodynamics and morphodynamics, a loos-coupling fluid-structure interaction (FSI) method (Borazjani et al., 2008) is employed. This method entails separating the problem into the fluid and sediment domains. We then separately solve the governing equations for flow and morphodynamics in each domain, while considering their interaction by applying appropriate boundary conditions at the sediment/water interface. Specifically, for solving the flow field equations, we specify boundary conditions at the bed, including bed location and velocity. However, to solve the bed change equation, we need velocity components and bed shear stress data from the flow domain to compute sediment fluxes (see Khosronejad et al., 2011). To solve the bed morphodynamics equation, we first determine $D_b$ and $E_b$ by solving Eqs. (6) and (7) for the suspended sediment concentration. Consequently, in each time-step, we follow a sequential process. In other words, we start by solving the Navier Stokes equations for the flow field, followed by resolving the equation for suspended sediment concentration, and lastly, solving the Exner-Polya equation to ascertain the updated bed elevations.

A significant hurdle in coupled numerical simulations of hydro- and morpho-dynamic processes in geophysical flows arises because the time scale of the flow is much shorter than the time scale of bed morphodynamics (Mercier et al., 2012). To address the computational challenges posed by the varying time scales of the flow and morphodynamic phases, we utilize a dual-time-stepping method in combination with a quasi-synchronization approach (see Khosronejad et al., 2014).

## 3. Computational details

To generate the training and validation dataset required for the ML model, we designed sixteen virtual meandering testbed rivers incorporating a wide range of bend shape and bend order, which has been suggested to affect the turbulent flow of meandering channels (Khosronejad et al. 2022). These testbeds are generated using a standard geometric model for the centerlines of meandering rivers, as follows:

$$\theta(s) = \theta_0 \sin\left(\frac{2\pi s}{\lambda}\right) + \theta_0^3 \left(J_s \cos\left(\frac{6\pi s}{\lambda}\right) - J_f \sin\left(\frac{6\pi s}{\lambda}\right)\right) \quad (9)$$

where $\theta$ is the local direction of the channel centerline, $s$ is the position along the centerline, $\lambda$ is bend wavelength, $\theta_0$ is the peak angular amplitude, $J_s$ is a skewness coefficient, and $J_f$ is a flatness coefficient. For parameters to construct the testbed rivers, the reader is refered to Khosronejad et al., 2022. The shapes of the designed meandering testbeds are shown in Figure 1. Moreover, the geometry and flow parameters (e.g., length, width, depth, sinuosity, and bulk velocity) of these sixteen testbed rivers are presented in Table 1.

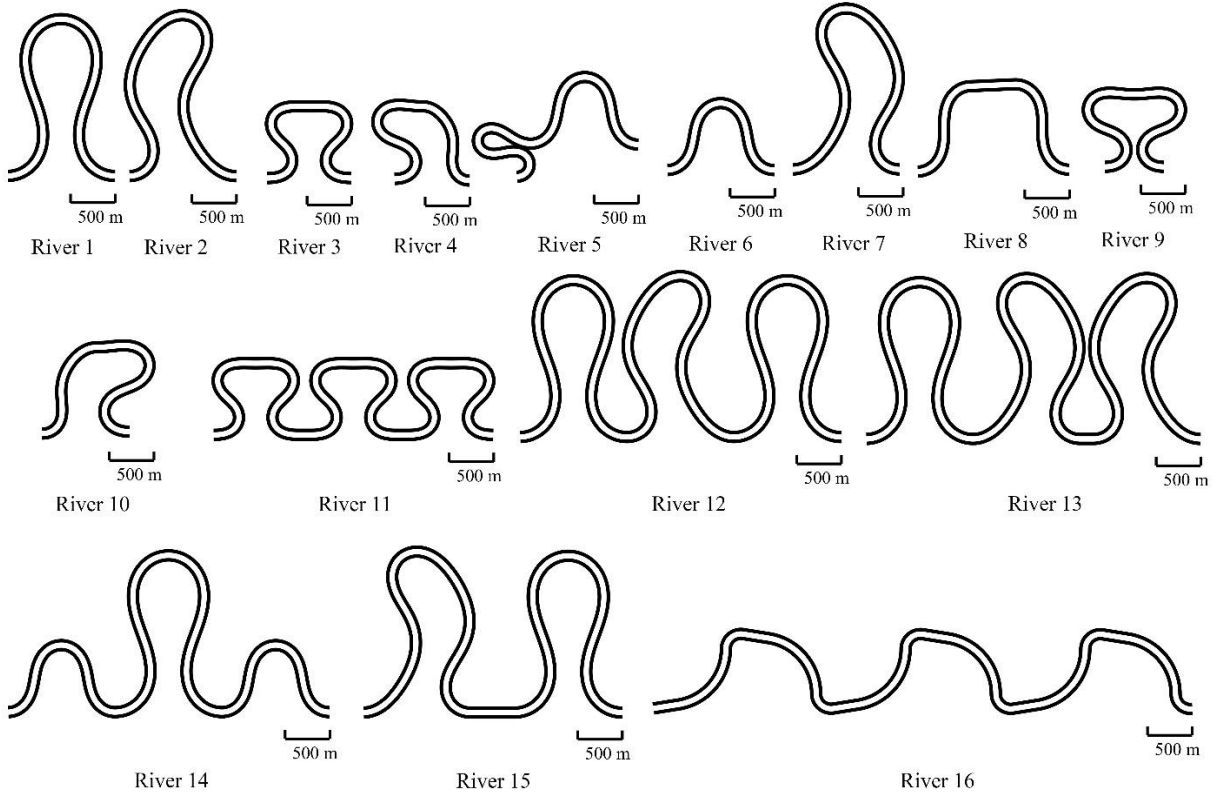

**Figure 1**: Platform geometry of the sixteen synthetic virtual meandering rivers used to generate training and validation dataset for the machine learning algorithm. The flow is from left to right.

**Table 1**: Geometrical and hydraulic characteristics of the sixteen riverbed rivers. L, W and H are the length, width, and depth of river. S is the slop. U is the bulk velocity of the flow.

|    | L (m) | W(m) | H(m) | S       | U(m/s) |
|----|-------|------|------|---------|--------|
| 1  | 4440  | 100  | 3.3  | 0.00016 | 1.75   |
| 2  | 4580  | 100  | 3.3  | 0.00016 | 1.75   |
| 3  | 2820  | 100  | 3.3  | 0.00016 | 1.75   |
| 4  | 2740  | 100  | 3.3  | 0.00016 | 1.75   |
| 5  | 3790  | 100  | 3.3  | 0.00016 | 1.75   |
| 6  | 2110  | 100  | 3.3  | 0.00016 | 1.75   |
| 7  | 4580  | 100  | 3.3  | 0.00016 | 1.75   |
| 8  | 3040  | 100  | 3.3  | 0.00016 | 1.75   |
| 9  | 3280  | 100  | 3.3  | 0.00016 | 1.75   |
| 10 | 3188  | 100  | 3.3  | 0.00016 | 1.75   |
| 11 | 8760  | 100  | 3.3  | 0.00016 | 1.75   |
| 12 | 13460 | 100  | 3.3  | 0.00016 | 1.75   |
| 13 | 13750 | 100  | 3.3  | 0.00016 | 1.75   |
| 14 | 8660  | 100  | 3.3  | 0.00016 | 1.75   |
| 15 | 9520  | 100  | 3.3  | 0.00016 | 1.75   |
| 16 | 8790  | 100  | 3.3  | 0.00016 | 1.75   |

To carry out the CFD and AI computations of turbulent flow and bed morphodynamics of the sixteen testbed rivers, the flow domains of the meandering rivers are discretized with structured background grid systems (Figure 2a), while the riverbeds are discretized with unstructured triangular grid systems and embedded in the flow domains (Figure 2b) -- as required by the CURVIB method. The separate grid systems of flow domain and riverbeds in the CURVIB approach enables the reconstruction of the mobile sediment bed during the simulation and the easy handling of the boundary condition at the interface between the water and riverbed sediment.

Table 2 shows the grid resolution and the time-steps used to simulate the turbulent flow and morphodynamics of the testbed rivers. The grid nodes of the background grid systems were spaced uniformly along the streamwise, spanwise, and vertical directions. Because the flow depths of the testbed rivers are much smaller than the river length and width, the grid resolutions in the vertical direction are smaller than in the longitudinal and spanwise directions. The unstructured triangular grid systems were also uniformly spaced along the channel bed. The grid systems were selected based on a series of grid sensitivity analyses and are adequate for resolving large-scale energetic coherent structures induced by the planform geometry of the meanders and the deformed geometry of their beds.

The time-step of the flow field computations, $\Delta t_f$, was selected to ensure that the Courant number was less than 1.0. The time-step for the morphodynamic calculations was set to $\Delta t_m = 500\Delta t_f$. The dual time-stepping desynchronization approach made the large-scale two-phase flow computations computational affordable in this study. And the $\Delta t_m$ was carefully selected to be sufficiently small to avoid numerical instability (Khosronejad et al. 2023).

The inlet boundary conditions of the computational domains were prescribed fully turbulent open channel flows calculated by separate precursor simulations. The precursor simulations were conducted in straight channels with rigid beds. The straight channels have cross-sections the same as the inlets of the testbed rivers and a length equal to double that of the river width, 2W. The precursor simulations used periodic boundary conditions in the streamwise direction and rigid lid on the free surface. Once the flow field is computationally converged, we stored the instantaneous flow field over a representative cross-plane for a sufficiently long sample as the inlet boundary condition of the testbed rivers. The Newman boundary condition was employed at the outlet of the testbed rivers for both the flow field and suspended sediment concentration, allowing the suspended sediment exit from the outlet cross plane. On the other hand, the total outflux of the suspended sediment concentration at the outlet was calculated at each time-step and recirculated into the inlet at the next time-step. The free surfaces of the testbed rivers were treated as sloping rigid lids with the slopes S in Table 1. The hydrodynamic effects of the solid surfaces (e.g., the mobile riverbed and side walls) on the flow domain were described by a wall model (Khosronejad et al. 2023) to reconstruct the velocity field at the first grid point off the solid boundaries since the viscous sublayer cannot be resolved directly.

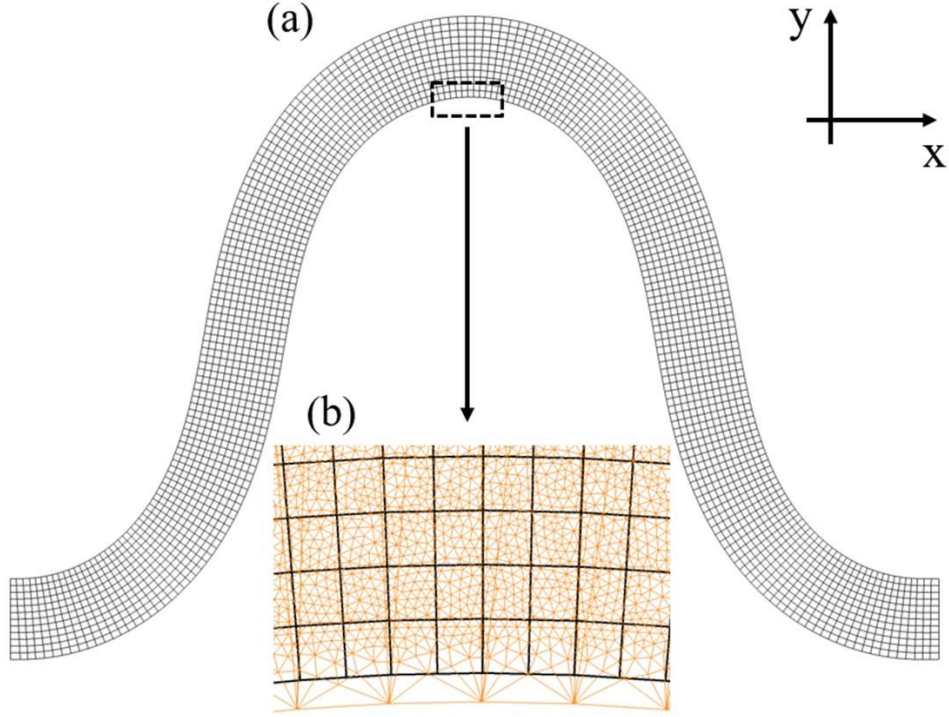

**Figure 2**: Schematic of the computational grid system to discretize a fluvial river in the context of CURVIB. (a) shows the structured grid system of the entire flow domain. (b) depicts the details of the flow domain and the unstructured triangular grid systems of the riverbed (orange lines).

**Table 2**: Computational grid systems and time-steps used in the coupled simulations of the sixteen rivers. $N_x$, $N_y$, and $N_z$ are the number of grid nodes to discretize the flow domain, which result in the grid resolution of $\Delta x$, $\Delta y$, and $\Delta z$ in longitudinal, spanwise, and vertical directions, respectively, and a total number of computational grid nodes $N_f$. The mobile sediment beds are discretized with $N_m$ triangular cells and a resolution of $\Delta s$. The time-steps of hydrodynamics and morphodynamics computations are $\Delta t_f$ and $\Delta t_m$, respectively.

| | $N_x \times N_y \times N_z$ | $\Delta x \times \Delta y \times \Delta z$ | $N_f \times 10^6$ | $\Delta s$ | $N_m \times 10^3$ | $\Delta t_f$ | $\Delta t_m$ |
|---|---|---|---|---|---|---|---|
| 1 | $10001 \times 225 \times 25$ | $0.44 \times 0.44 \times 0.14$ | 56.25 | 0.98 | 449.0 | 0.06 | 30 |
| 2 | $10013 \times 225 \times 25$ | $0.46 \times 0.44 \times 0.14$ | 56.32 | 1.00 | 450.6 | 0.06 | 30 |
| 3 | $6101 \times 225 \times 25$ | $0.46 \times 0.44 \times 0.14$ | 34.32 | 0.98 | 274.54 | 0.06 | 30 |
| 4 | $6001 \times 225 \times 25$ | $0.46 \times 0.44 \times 0.14$ | 33.76 | 0.98 | 271.22 | 0.06 | 30 |
| 5 | $8201 \times 225 \times 25$ | $0.46 \times 0.44 \times 0.14$ | 46.13 | 0.98 | 274.54 | 0.06 | 30 |
| 6 | $5001 \times 225 \times 25$ | $0.42 \times 0.44 \times 0.14$ | 28.13 | 0.95 | 225.0 | 0.06 | 30 |
| 7 | $10013 \times 225 \times 25$ | $0.46 \times 0.44 \times 0.14$ | 56.32 | 1.00 | 450.6 | 0.06 | 30 |
| 8 | $6753 \times 225 \times 25$ | $0.45 \times 0.44 \times 0.14$ | 37.99 | 0.99 | 303.88 | 0.06 | 30 |
| 9 | $7101 \times 225 \times 25$ | $0.46 \times 0.44 \times 0.14$ | 39.94 | 0.99 | 319.54 | 0.06 | 30 |
| 10 | $6901 \times 225 \times 25$ | $0.46 \times 0.44 \times 0.14$ | 38.82 | 0.99 | 310.99 | 0.06 | 30 |
| 11 | $19001 \times 225 \times 25$ | $0.46 \times 0.44 \times 0.14$ | 106.88 | 1.00 | 855.03 | 0.06 | 30 |
| 12 | $26921 \times 225 \times 25$ | $0.46 \times 0.44 \times 0.14$ | 106.88 | 1.00 | 1211.14 | 0.06 | 30 |
| 13 | $27481 \times 225 \times 25$ | $0.50 \times 0.44 \times 0.14$ | 154.58 | 1.04 | 1236.64 | 0.06 | 30 |
| 14 | $19001 \times 225 \times 25$ | $0.46 \times 0.44 \times 0.14$ | 106.88 | 1.00 | 855.03 | 0.06 | 30 |

| 15 | 19001 × 225 × 25 | 0.50 × 0.44 × 0.14 | 106.88 | 1.04 | 855.03 | 0.06 | 30 |
| 16 | 19001 × 225 × 25 | 0.46 × 0.44 × 0.14 | 106.88 | 1.04 | 855.03 | 0.06 | 30 |

The simulations were carried out using 80 to 420 Intel Xeon 3.3GHz processors for the testbed rivers with different number of grid nodes. We first carried out hydrodynamic simulations for the testbed rivers with rigid flat bed for at least two flow-through time to obtain a fully developed turbulent flow field. The flow-through time is the time for a water particle to travel from the inlet to the outlet. Then, the coupled flow and morphodynamic simulations were executed until the mobile riverbeds reached dynamic equilibrium. The criterion of dynamic equilibrium is defined as the maximum local bed change for 10 successive morphodynamics time steps was less than one percent of the mean flow depth. Subsequently, the coupled flow and morphodynamic simulations were continued to perform the time-averaging until the flow fields were statistically converged. The hydrodynamic simulations cost approximately 720 to 10,000 CPU hrs, and the coupled flow and morphodynamic simulations required approximately 11,000 to 450,000 CPU hrs for the shortest and longest rivers, with lowest and highest number of computational grid nodes, respectively.

## 4. Machine Learning Algorithm

Past studies (e.g., Santoni et al., 2023; Zhang et al., 2023, 2024; Zhang, Flora, et al., 2022b; Zhang, Santoni, et al., 2022) demonstrated the potential of CNNAE to learn the non-linear relations between snapshots of turbulent flow field and time-averaged flow fields at high Reynolds. Since the computational cost of the coupled flow and morphodynamic simulations is at least one order of magnitude greater than the hydrodynamic simulations, a first logical thought would involve extending the application of the CNNAE to predicting the time-averaged shear stress and equilibrium bed elevation based on their instantaneous snapshots taken from the very early time steps of the coupled flow and morphodynamic simulation results. This section describes the architecture and the workflow of the CNNAE we implemented to do so.

### 4.1 Convolutional neural networks autoencoder

As seen in the schematic of Figure 3, the CNNAE uses instantaneous flow field results from coupled LES-morphodynamics and the geometry parameters of the testbed river as inputs to predict the time-averaged and/or equilibrium morphodynamical variables, such as the shear stress and the bed elevation. We note that the bed shear stress and the bed elevation results consist of 2D dataset projected on the riverbed surface. The CNNAE consists of (*i*) an encoder which maps the high-dimensional input data to a low-dimensional latent space, and (*ii*) a decoder which reconstruct the high-dimensional output data from the low-dimensional latent space. The encoder consists of four 2D convolutional layers (Conv2D), and the decoder consists of three 2D transposed convolutional layers (ConvT2D) followed by a 2D convolutional layer, as presented in Table 3. The convolutional layers are defined as:

$$y_n = \sum_m k_{m,n} * x_m + b_n \qquad (10)$$

where $x_m$ and $y_n$ denote the $m$th input channel and the $n$th output channel of the layer, respectively, $k_{m,n}$ is the convolutional kernel, $*$ denotes the convolution operator, and $b_n$ is the bias of the $n$th output channel. In each layer, $m \times n$ convolutional kernels traverse through $m$ channels of the layer's input to extract $n$ feature maps. Traverse step sizes (stride) greater than one are used here to down sample the feature maps. As described in Table 1, the length of the rivers is significantly greater than their width and water depth. As a result, the dimension of the rivers in streamwise direction is much larger than in spanwise direction, the dimension of the kernel size and the stride in streamwise direction are accordingly designed larger than in spanwise direction. The stride in the streamwise direction is determined by a systematical study of the effect of receptive field size on the validation MSE. The receptive field is the product of the stride size of all the layers in encoder, representing the input area that each element in the bottleneck latent space can "see". Since the main features of the meandering river have large scale along the streamwise direction, a larger receptive field might be needed to capture those features. As seen in Figure 4, the validation MSE significantly decrease as the receptive field size increase in the streamwise direction, and plateaued above receptive field size of 64. Therefore, the receptive field size of 80 is used in our study. A padding is applied to each layer's input to fine tune the output dimension. After each convolutional layer, a leaky ReLU function σ is applied to bring the nonlinearity to the model, which is given by (Maas et al. 2013)

$$\sigma(y) = \begin{cases} y, & y > 0 \\ 0.01y, & y < 0 \end{cases} \quad (11)$$

The transposed convolutional layers are the adjoint operation of convolution, which can reconstruct the high-dimensional data from the lower-dimensional feature maps. Accordingly, the input and output dimensions are the reverse of the corresponding convolutional layers.

During the training process, the learnable parameters of convolutional kernel $k_{m,n}$ and bias $b_n$ are updated using a backpropagation algorithm to minimize the objective function, given by the mean square error (MSE) of the discrepancy between the predicted results and the ground truth, as follows:

$$Loss = MSE(\psi_{CNN} - \psi_{LES}) \quad (12)$$

where $\psi_{CNN}$ and $\psi_{LES}$ are the CNNAE outputs and the coupled LES-morphodynamics time-averaged results, respectively. The optimizer employed in the proposed algorithm is Adam (Kingma and Ba 2014). The learning rate of the training process had an initial value of 0.001 with a decay rate of 0.7 in a step size of 400 training epochs. A coarse grid search was conducted over the hyperparameter space to determine the optimal number of layers, channels, kernel sizes, and strides by maximizing the generality and performance of the trained model.

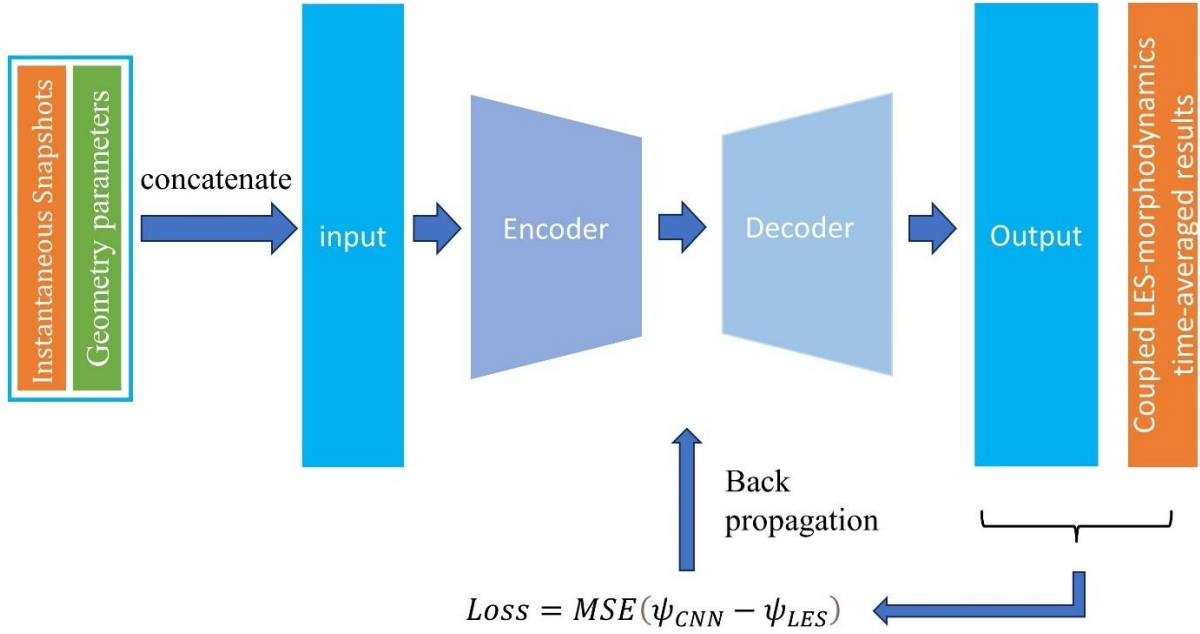

**Figure 3:** The schematic of the CNNAE.

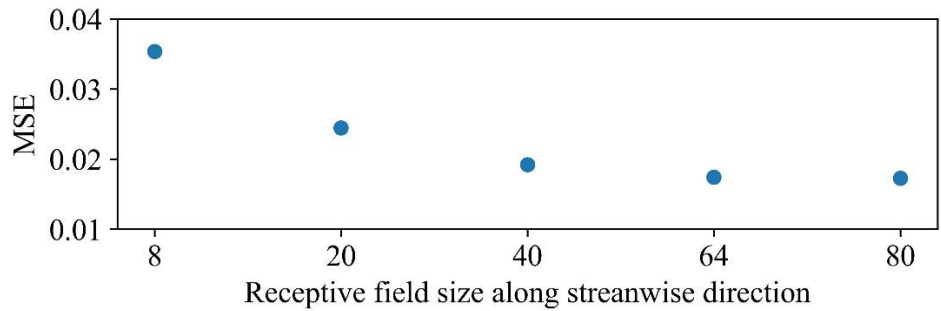

**Figure 4:** Validation MSE of CNNAEs with different receptive field size along streamwise direction.

**Table 3**: Architecture of the CNNAE.

|   | **Layer** | **Channel** | **Kernel** | **Stride** | **Padding** |
|---|---|---|---|---|---|
| 1 | Conv2d | (input, 8) | (7,3) | (1, 1) | (3, 1) |
| 2 | Conv2d | (8, 16) | (7, 3) | (4, 2) | (3, 1) |
| 3 | Conv2d | (16, 32) | (7, 3) | (4, 2) | (3, 1) |
| 4 | Conv2d | (32, 64) | (9, 3) | (5, 2) | (4, 1) |
| 5 | ConvT2d | (64, 32) | (9, 4) | (5, 2) | (2, 1) |
| 6 | ConvT2d | (32, 16) | (6, 4) | (4, 2) | (1, 1) |
| 7 | ConvT2d | (16, 8) | (6, 4) | (4, 2) | (1, 1) |

| 8 | Conv2d | (8, 1) | (7, 3) | (1, 1) | (3, 1) |

**4.2 Training and prediction workflow**
In our previous studies, we validated the ability of CNNAE in recognizing the patterns in instantaneous turbulent flow field to predict the mean velocity field and turbulence kinetic energy on a variety of test cases, such as meandering rivers with bridge piers (Zhang, et al., 2022), wind-wave interaction over oceanic waves (Zhang et al. 2023), wake flow of wind turbines (Santoni et al., 2023; Zhang, et al., 2022), and wake flow of marine hydrokinetic turbines in meandering rivers (Zhang et al. 2024). The proposed algorithms showed satisfactory accuracy in reconstructing turbulence statistics while using only a small portion of the high-fidelity simulation results, i.e., several velocity field snapshots. Therefore, we expect to apply the same idea in the prediction of river bed morphodynamics and bed shear stress distribution to reduce the computational cost of the simulations. Specifically, we intend to use the instantaneous results of the first few time-steps of the coupled flow and morphodynamics simulation to predict the time-averaged bed shear stress distribution on river bed and bed elevation at equilibrium. However, the prediction of river bed evolution involves two challenges: (*i*) the bed elevation, $Z_b$, is very small and near machine zero at the first few time-steps of the coupled flow and morphodynamics simulation. For that, it cannot provide any significant pattern for the CNNAE to recognize, and (*ii*) the time-averaged bed shear stress distribution not only depends on its instantaneous snapshots but also is affected by the bed elevation. To handle these challenges, we assume that (*i*) the bed elevation change, $\Delta Z_b$, and its patterns contains sufficient information for the CNNAE to learn from, and (*ii*) appropriate geometrical parameters of the river could help fine-tune the effects of the bed deformation, because both the shear stress distribution and bed elevation change are shown to be related to the geometry of the meander bends (Khosronejad et al. 2023).

Instead of cartesian coordinates *x* and *y*, we use the curvilinear coordinates (e.g., the streamwise distance from the inlet *s*, and the spanwise distance from the centerline of river *n*) along with the local curvatures of river bends *c* as the geometry parameters to fine-tune the CNNAE prediction. The coordinate parameters are stored on the structured background grid system. The schematic of the curvilinear coordinates is shown in Figure 5a. The local curvature on each grid node is defined by the reciprocal of the radius of inscribed circle of the streamwise coordinate line. The curvilinear coordinates *s* and *n* are normalized by the length *L* and width *W* of the river, respectively, to serve as the input vectors to the CNNAE. The schematics of the input geometrical parameters of CNNAE are shown in Figure 5b to d.

Figure 6 illustrates the workflow of the proposed algorithm. The hydrodynamic simulation is first conducted using coupled LES-morphodynamics to produce the fully developed turbulent flows in testbed rivers. During this simulation, the river beds are rigid so that the bed elevation $Z_b = 0$. Then, the coupled flow and morphodynamics simulation starts and run for few time-steps to produce the instantaneous shear stress distribution $\tau_b$ and change of bed elevation $\Delta Z_b$. The bed shear stress and bed elevation are resolved on the unstructured triangle mesh system covering the riverbed. For that, the information is projected on the structured background grid system where it is processed by the CNNAE. The instantaneous $\tau_b$ and $\Delta Z_b$ data projected on the structured grid system are normalized to the range of 0 to 1 and -1 to 1, respectively, and are concatenated with the geometry parameters $s/L$, $n/W$, and $c$ to render them all ready as the inputs to the CNNAE. The coupled flow and morphodynamics simulation continue until the bed elevation reaches equilibrium and then start generating samples and time-averaging the data to produce the time-averaged bed shear stress distribution and bed elevation as the training and validation targets of the CNNAE. Two CNNAE algorithms are trained separately for the prediction of $\tau_b$ and $\Delta Z_b$. Rivers 1 to 5 were used to train the CNNAE models, and rivers 6 to 16 were used to validate the trained models. The training dataset consist of instantaneous snapshots from the first 30 time-steps of the coupled flow and morphodynamics simulation with an interval of 100 time-steps for each river. While the validation dataset includes the first 5 time-steps for each river. The trainings were performed on an Inter Haswell CPUs and spent 20 CPU hrs to run 10,000 epochs to converge.

To validate the effectiveness of the geometry parameters introduced as inputs to CNNAE, we conducted a series of analysis to compare the effects of different combination of geometry parameters, as inputs, on the prediction accuracy. Eight algorithms were trained with no geometry parameter (Ng), single geometry parameter (s, n, c), double geometry parameters (sn, sc, nc), and all the three geometry parameters (snc) as the input of CNNAE. The (validation) mean square errors (MSE) of the trained CNNAE algorithm for $\tau_b$ and $\Delta Z_b$ are plotted in Figures 7a and 7b, respectively. As seen, the results demonstrate that, when the geometry parameters are introduced as the inputs, they evidently reduced the validation errors, namely, improving the generality of the trained machine learning algorithms.

In a recent study, Zhang et al., 2022 reported that using more instantaneous snapshots from different time-steps as input vector can eliminate the negative effects of large-scale turbulence structures on the accuracy of the CNNAE predictions. Herein, we also conducted a series of tests to assess the effects of the number of inputs, i.e., the number of instantaneous snapshots. Five cases with 1 to 5 consecutive instantaneous snapshots as the inputs of CNNAE with interval of 100 time-steps were considered. The CNNAE algorithms were trained separately for each case. The (validation) MSE of the trained CNNAE for $\tau_b$ and $\Delta Z_b$ are shown in Figure 8a and 8b, respectively. The results of this analysis showed that the number of instantaneous snapshots as inputs has near zero impact on the accuracy of the trained CNNAE model. Therefore, only one snapshot taken from the beginning of the coupled flow and morphodynamics simulation (i.e., the first time-step) deemed sufficient for the prediction.

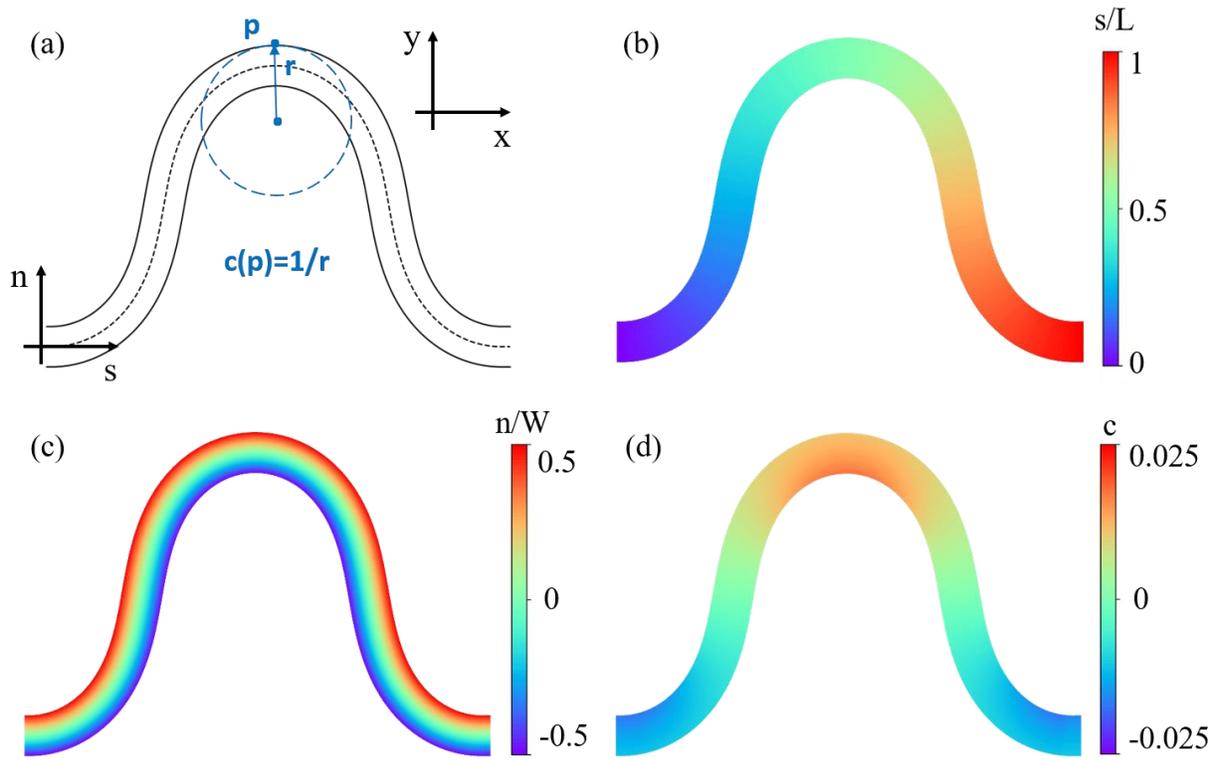

**Figure 5**: Schematic of (a) the curvilinear coordinate system and the calculation of local curvature, (b) the streamwise coordinate *s* normalized by the river length, (c) the spanwise coordinate *n* normalized by the river width, (d) the local curvature.

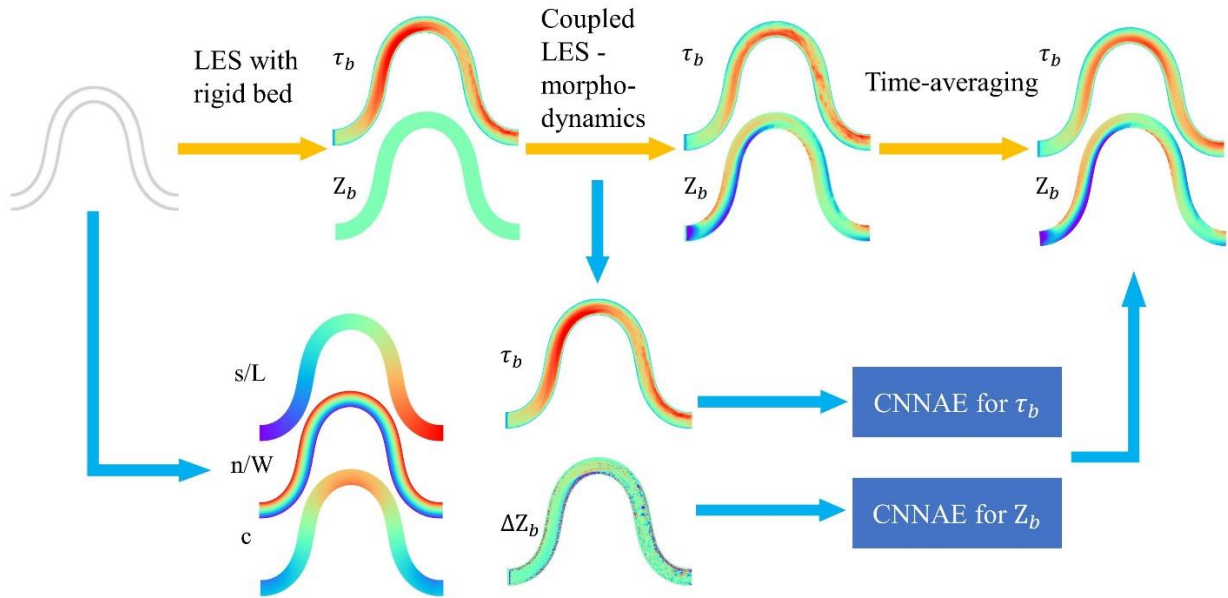

**Figure 6**: The workflow of the proposed CNNAE prediction process.

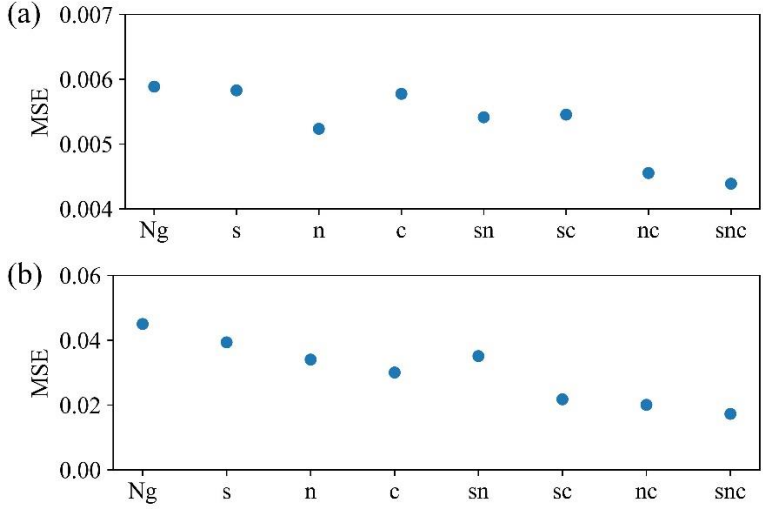

**Figure 7**: Validation MSE of eight combinations of geometry parameters. (a) the CNNAE for shear stress distribution prediction. (b) the CNNAE for bed elevation prediction. Ng means no geometry inputs. $s, n, c$ stand for streamwise coordinate, spanwise coordinate, and local curvature, respectively.

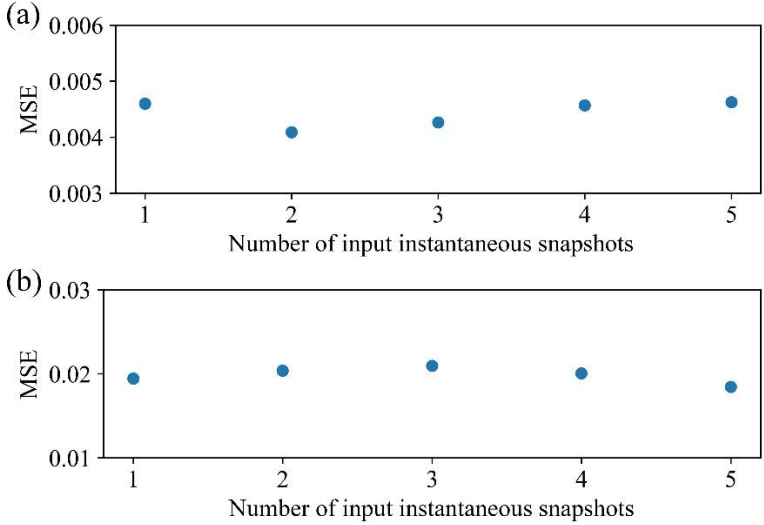

**Figure 8**: Validation MSE of different number of input instantaneous snapshots. (a) the CNNAE for shear stress distribution prediction. (b) the CNNAE for bed elevation prediction.

## 5. Results and discussion

In this section, we present the predicted shear stress distribution and bed elevation of the validation rivers 6 to 16 by the trained CNNAE models, and compared the predictions with the high-fidelity coupled LES-morphodynamics results to evaluate the performance of the proposed CNNAE algorithm.

### 5.1 Prediction of bed shear stress distribution at equilibrium

The bed shear stress distribution ($\tau_b$) is expressed in terms of dimensionless Shields parameter:

$$\theta = \frac{\tau_b}{(\rho_s - \rho)gD_{50}} \tag{13}$$

where $\rho_s$ is the density of the sediment material (=2,650 kg/m$^3$), $g$ is the gravitational acceleration, and $D_{50}$ is the median grain size of the bed material (=0.001 m). Figure 9 presents the instantaneous Shield parameter distributions at the 100th time-step of the coupled simulations, which construct the input vector to the CNNAE for shear stress distribution prediction. As seen in this figure, the imprints of the large-scale turbulence coherent structures can be clearly observed in the high Shield parameter regions marking the time variation of bed shear stress distribution. The CNNAE algorithm is expected to time-average such variations. On the other hand, since the input snapshots are from the beginning of the bed deformation when the riverbeds are nearly flat, the bed shear stress distribution is different from the deformed riverbeds at dynamic equilibrium. For example, lets focus on the time-averaged Shield parameter distribution over deformed riverbeds at equilibrium, as seen in Figure 11. The Shield parameter near the inner-bank of the bends and upstream the first bends of the meanders are higher than those observed on the deformed riverbeds at equilibrium (e.g., see rivers 6 to 10 of Figure 11). The trained CNNAE is also expected to amend such discrepancies in shear stress distribution of the flat and deformed riverbeds.

Figure 10 and 11 depict the CNNAE predictions and the coupled LES-morphodynamics results of the time-averaged Shield parameter distributions. In Figure 10, the imprints of the turbulence coherent structures are smoothed out, while the main features of the bed shear stress distribution in meandering rivers are captured. Such features, which are also reported in Khosronejad et al., 2023, include: (*i*) the high bed shear stress regions are aligned with the curves of the meanders and located close to the inner banks at the apexes; (*ii*) immediately downstream from the apexes, the high bed shear stress regions begin to move away from the inner-bank curves and positioned closer to the centerlines and outer-bank curves of the rivers. Surprisingly, the CNNAE correctly predicted the shear stress distribution over the deformed riverbed at equilibrium – despite that the input vectors to the trained CNNAE included the bed shear stress data over flat riverbeds. We argue that this adjustment implemented by the trained CNNAE could be attributed to the geometry parameters introduced in the input vectors.

Figure 12 presents the prediction error of the CNNAE. For a quantitative comparison, we plot in Figure 13 the longitudinal profiles of the instantaneous input vectors, the CNNAE predictions, and the coupled LES-morphodynamics results along two streamlines 0.2W away from the outer and inner bank, respectively. In the single bend rivers 6 to 10, the CNNAE has an impressive accuracy for most of the regions of the rivers, while it slightly underestimated the shear stress near the riverbanks. In river 11, the CNNAE evidently underestimated the shear stress in most of the region, however, it corrected the distribution of the high shear stress region at the straight channels downstream of the apex. More specifically, in the instantaneous input vector of the bed shear stress, the high shear stress region shifts to the outer-bank immediately downstream the apex, however, in the predicted target vector and the time-averaged couple LES-morphodynamics results, the high shear stress regions are located around the centerline of the river. This correction can also be observed in rivers 13 and 15, at the region immediately downstream the apexes of the bend.

Overall, these prediction results show that the implemented CNNAE has a great potential to generate 3D realizations of the bed shear stress distribution in high-Reynolds riverine flows. Taking advantage of the proposed AI algorithm could help reducing the cost of mean bed shear stress distribution significantly, i.e., by 98%, compared to the coupled LES-morphodynamics model.

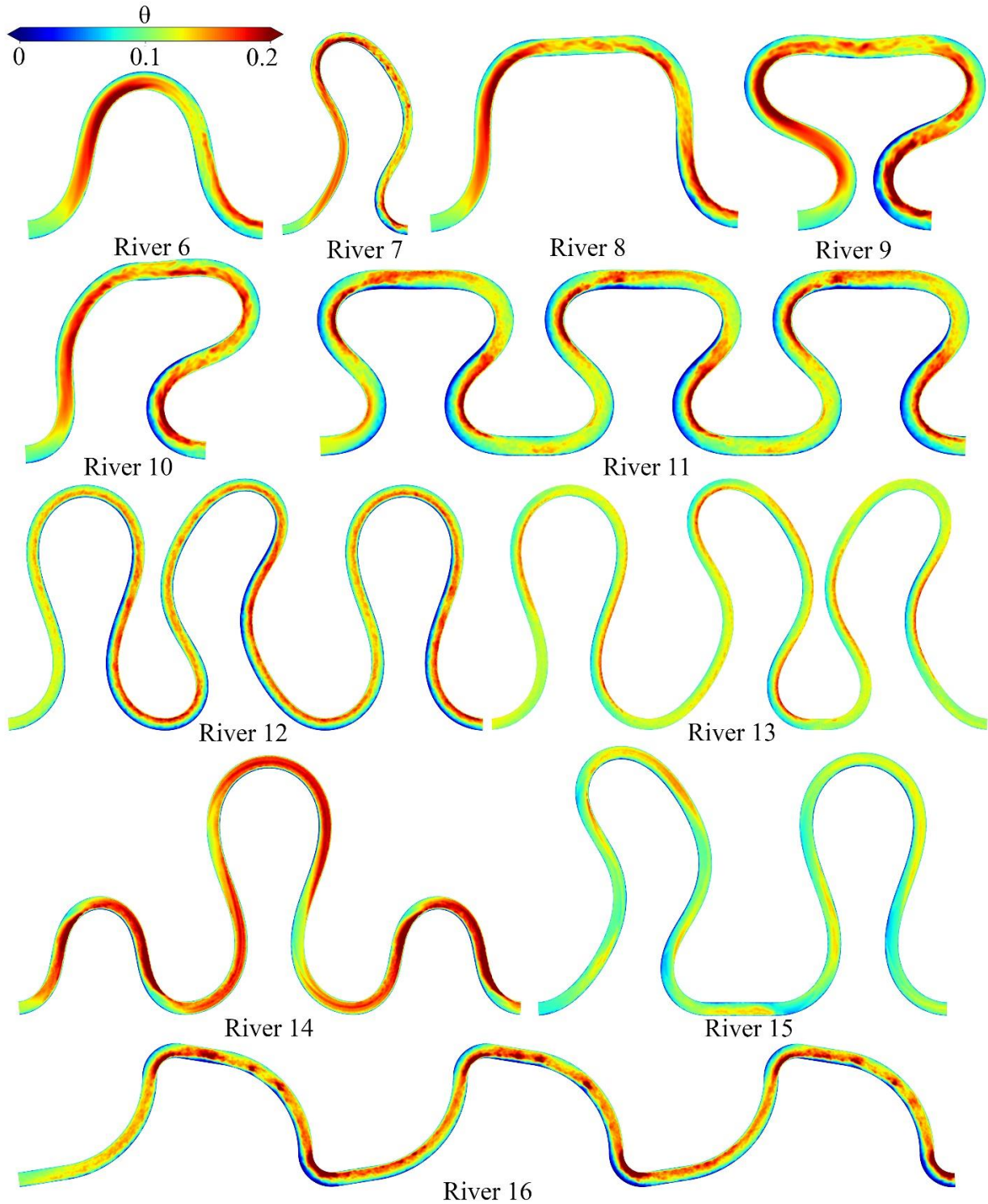

**Figure 9**: Instantaneous Shield parameter distribution on riverbeds at time-step 100 of the coupled LES-morphodynamics simulation. These snapshots constitute the input vector to the trained CNNAE to predict mean bed shear stress distributions.

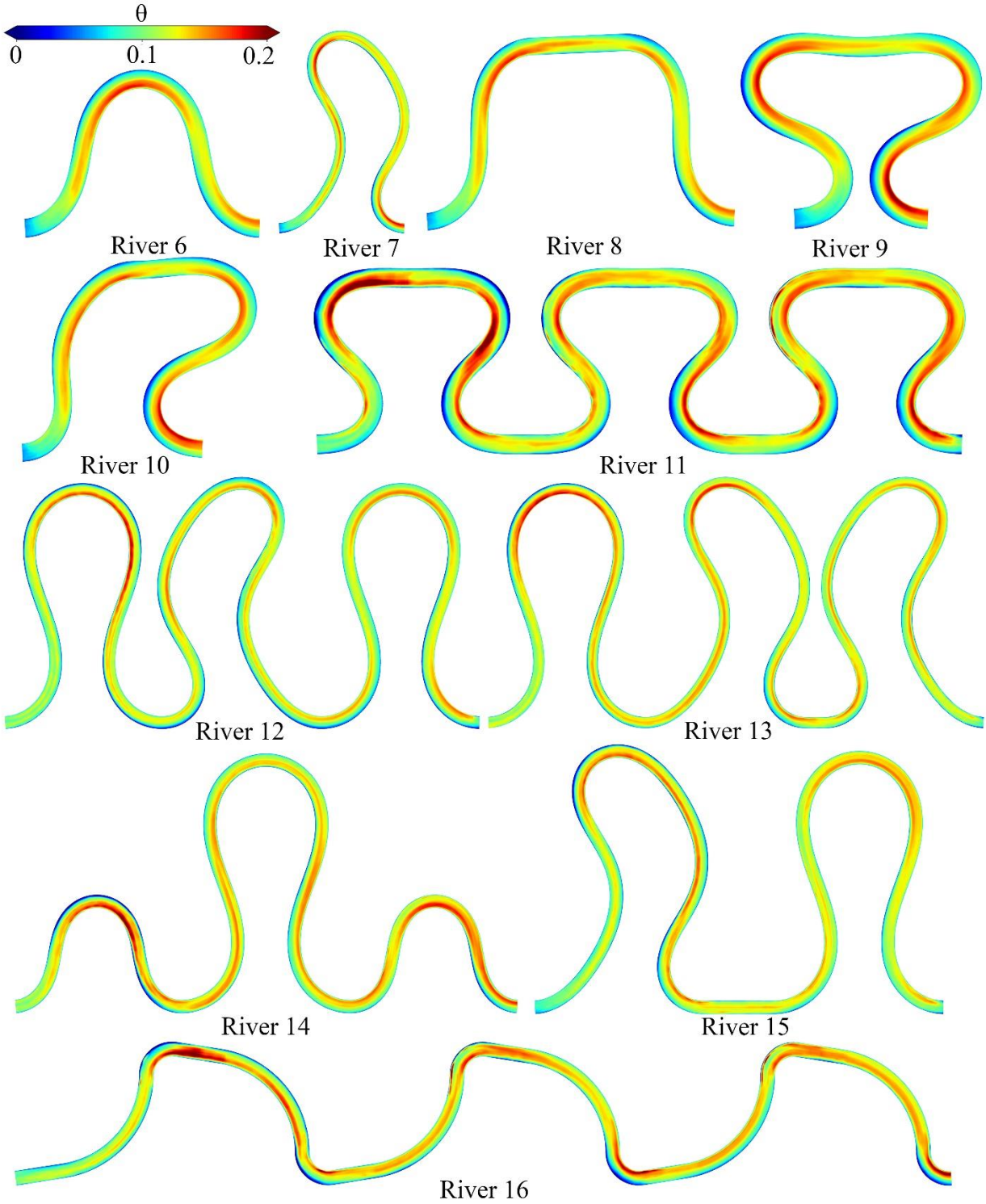

**Figure 10**: CNNAE prediction results of the time-averaged Shield parameter distribution.

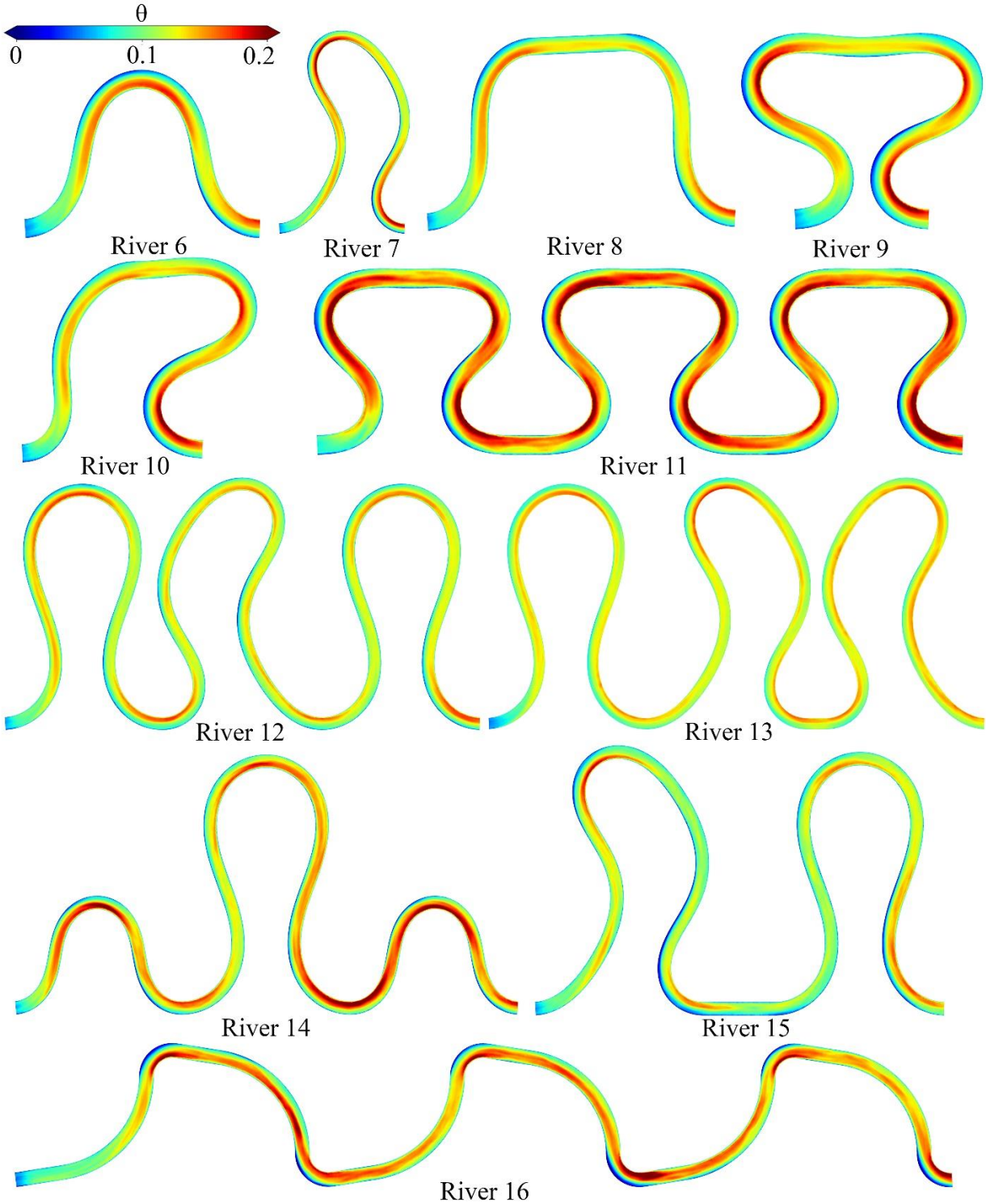

**Figure 11**: Coupled LES-morphodynamics simulation results for the time-averaged Shield parameter distribution.

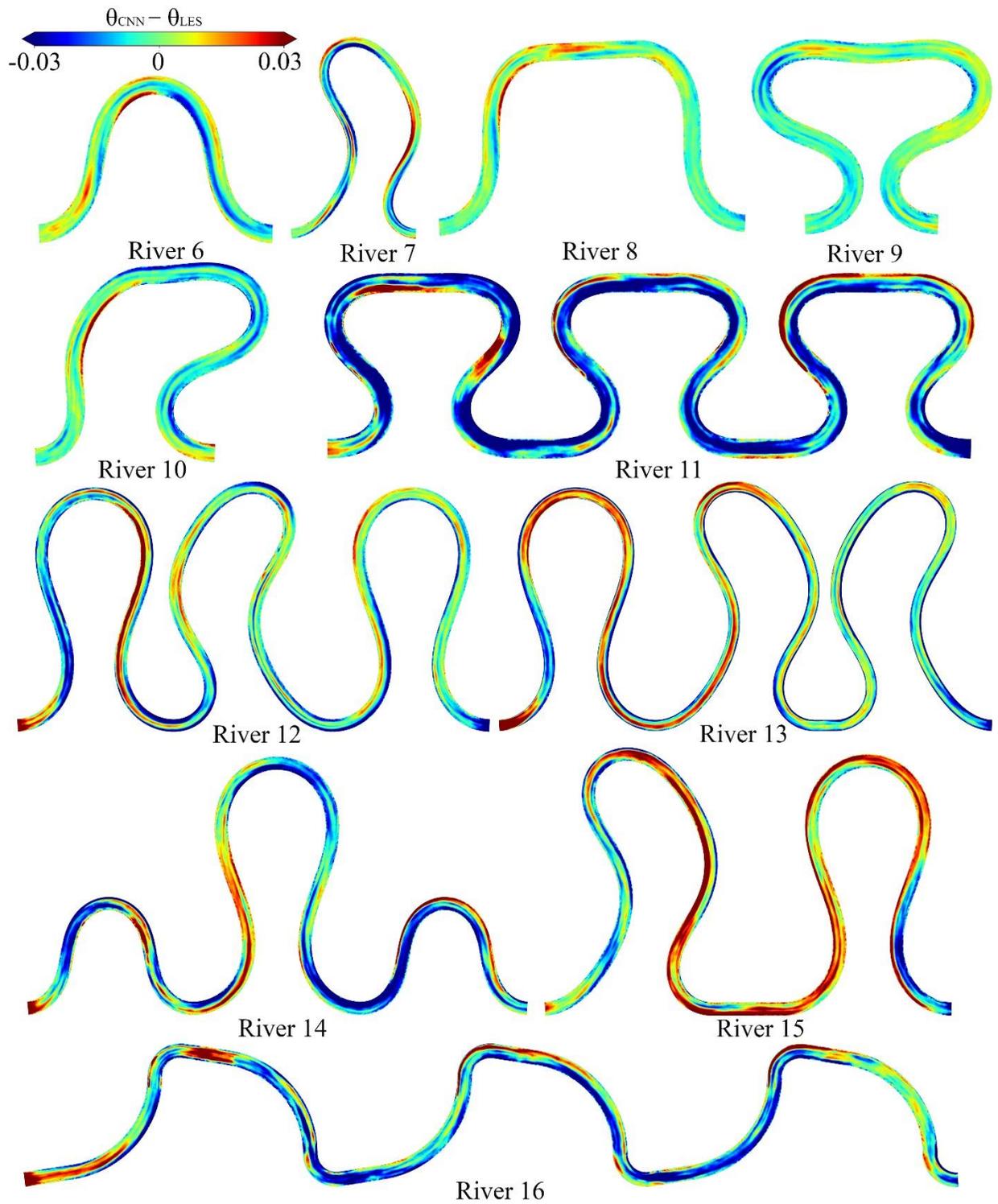

**Figure 12**: The prediction error of the Shield parameter distribution.

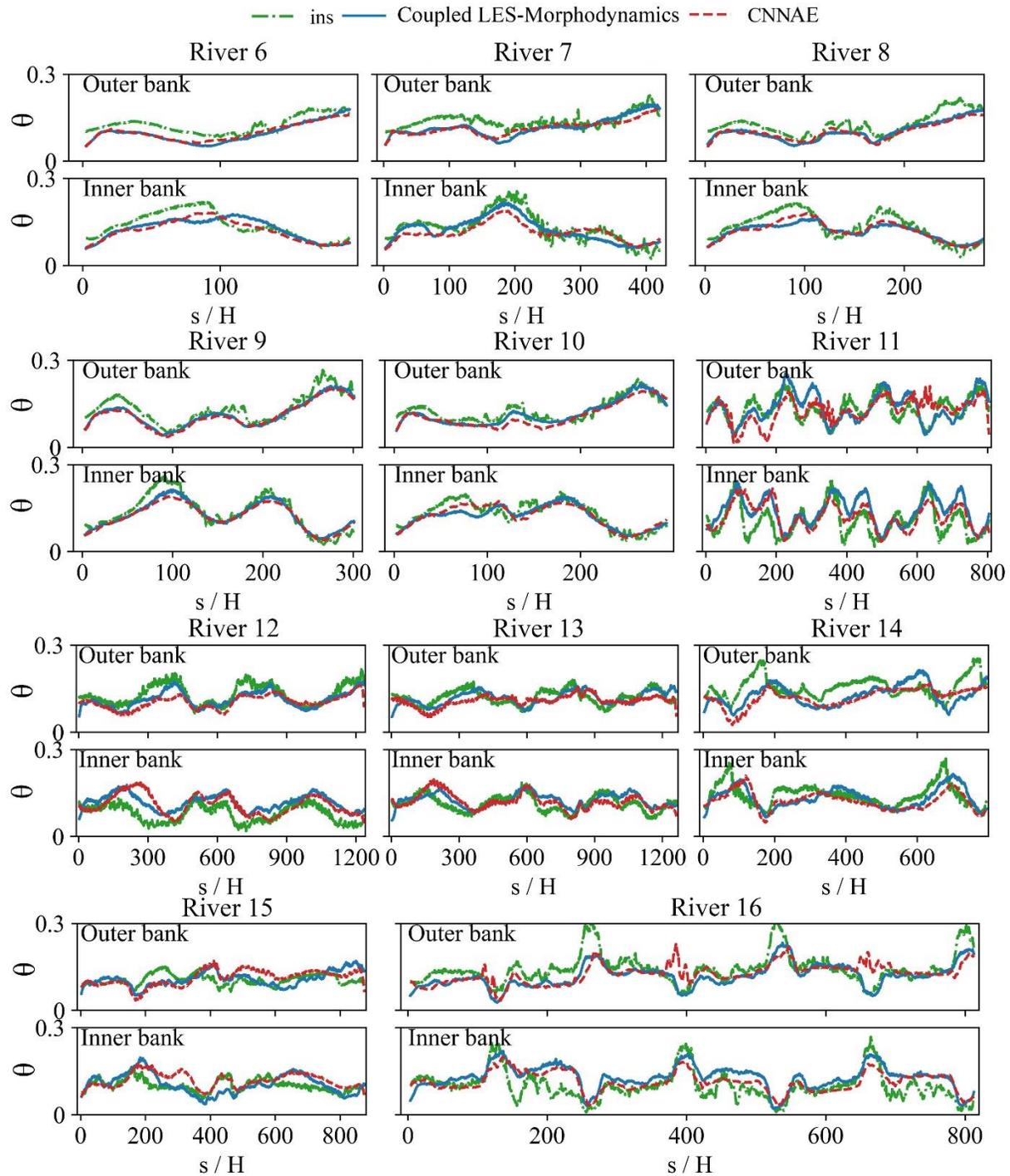

**Figure 13**: The profiles of the Shield parameter along the "s" vector in the streamwise direction. Profiles are extracted from streamwise "s" vectors located 0.2W from the outer and inner banks. Green dotted-dashed, blue solid, and red dashed lines represent the instantaneous results of the coupled LES-morphodynamics model, time-averaged results of the coupled LES-morphodynamics model, and equilibrium results of the CNNAE algorithm.

## 5.2 Prediction of bed topography of rivers at dynamic equilibrium

Figure 14 presents the input vector of the bed elevation prediction for validation of the trained CNNAE. They are instantaneous bed change $\Delta Z$ at the 100th time-step of the coupled simulation non-dimensionalized by the water depth $H$. Unlike the input vectors for the shear stress distribution prediction, which are somewhat similar with the target vector of time-averaged bed shear stress distribution, the input vectors for bed change does not have any recognizable pattern that replicating the target vector of the equilibrium bed topography.

Figure 15 and 16 present the CNNAE and the coupled LES-morphodynamics (time-averaged) bed elevation normalized with the water depth H, respectively. Impressively, the trained CNNAE successfully reconstructed the bed elevations at equilibrium using the instantaneous bed change data. As seen, the scour and deposition regions at the outer and inner banks were correctly captured. Further, the main features of the sediment transport and bed deformation in large-sale meandering rivers are captured qualitatively well with the CNNAE. Some of such features include the marked scour regions at the outer bends of the apexes and the sediment deposition regions near the inner banks. For more details of such dominant features concerning the sediment transport in large-sale meandering rivers, we refer the reader to Khosronejad et al., (2023).

Figures 17 and 18 depict the prediction error of the CNNAE equilibrium bed elevation results relative to the time-averaged results of the coupled LES-morphodyanics model. These bed elevation profiles are recorded along the streamwise lines 0.2W away from the outer and inner banks to provide more details evaluating the prediction performance of CNNAE. In the rivers with a single bend – rivers 6 to 10, the prediction of scour and deposition regions, and bed elevation in general, were mostly accurate. However, in the region upstream of the apex in river 6, the scour and deposition are underestimated, and the scour region at the outer bank of river 10 was not captured as expected. The accuracy of predictions in rivers 11 to 16 were relatively lower than the rivers with a single bend. For instance, the scour and deposition regions in rivers 12 to 15 were expected to form downstream of the apexes, however, they are captured further upstream the apexes in the CNNAE-predicted bed elevation contours. Although the bed elevation prediction results of the CNNAE plotted in Figure 17 have some discrepancies, the regions of the scour and deposition are captured correctly.

Now we focus our attention on the computational cost of the predictions. The hydrodynamic simulations for each river cost approximately 720-10,000 CPU hrs, while the coupled flow and morphodynamic simulations required approximately 11,000-450,000 CPU hrs for the rivers with lowest and highest number of grid nodes, respectively. On the other hand, the proposed CNNAE algorithm need only 20 hrs to train and about 1 min to infer the new cases, considering the cost of hydrodynamic simulations and the first 100 time-steps of the coupled simulations to generate the inputs of the CNNAE. Thus, the total cost of the proposed AI algorithm to infer the bed shear stress distribution and equilibrium bed elevation of a new river is less than 2% of the high-fidelity coupled LES-morphodynamics simulations. The high efficiency of this method could help to save a huge amount of computational cost in the systematical study of bed changes of meandering rivers like in (Khosronejad et al. 2023).

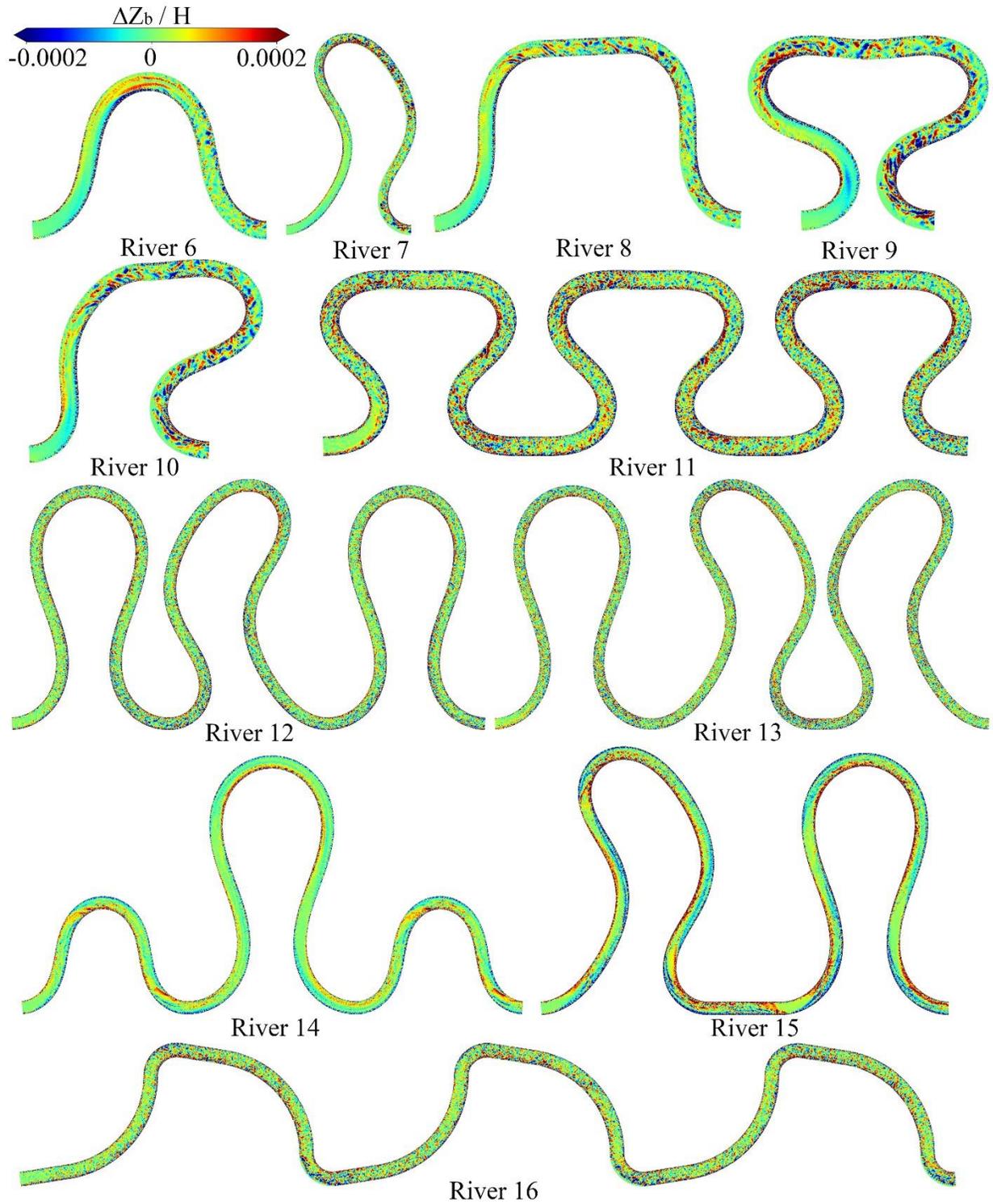

**Figure 14**: Instantaneous bed elevation change data, non-dimensionalized by the river depth H, obtained from the coupled LES-morphodynamics model at time-step 100 -- nearly beginning of the simulation. This dataset is a part of the input vector to the CNNAE to predict the bed elevation of the river at equilibrium.

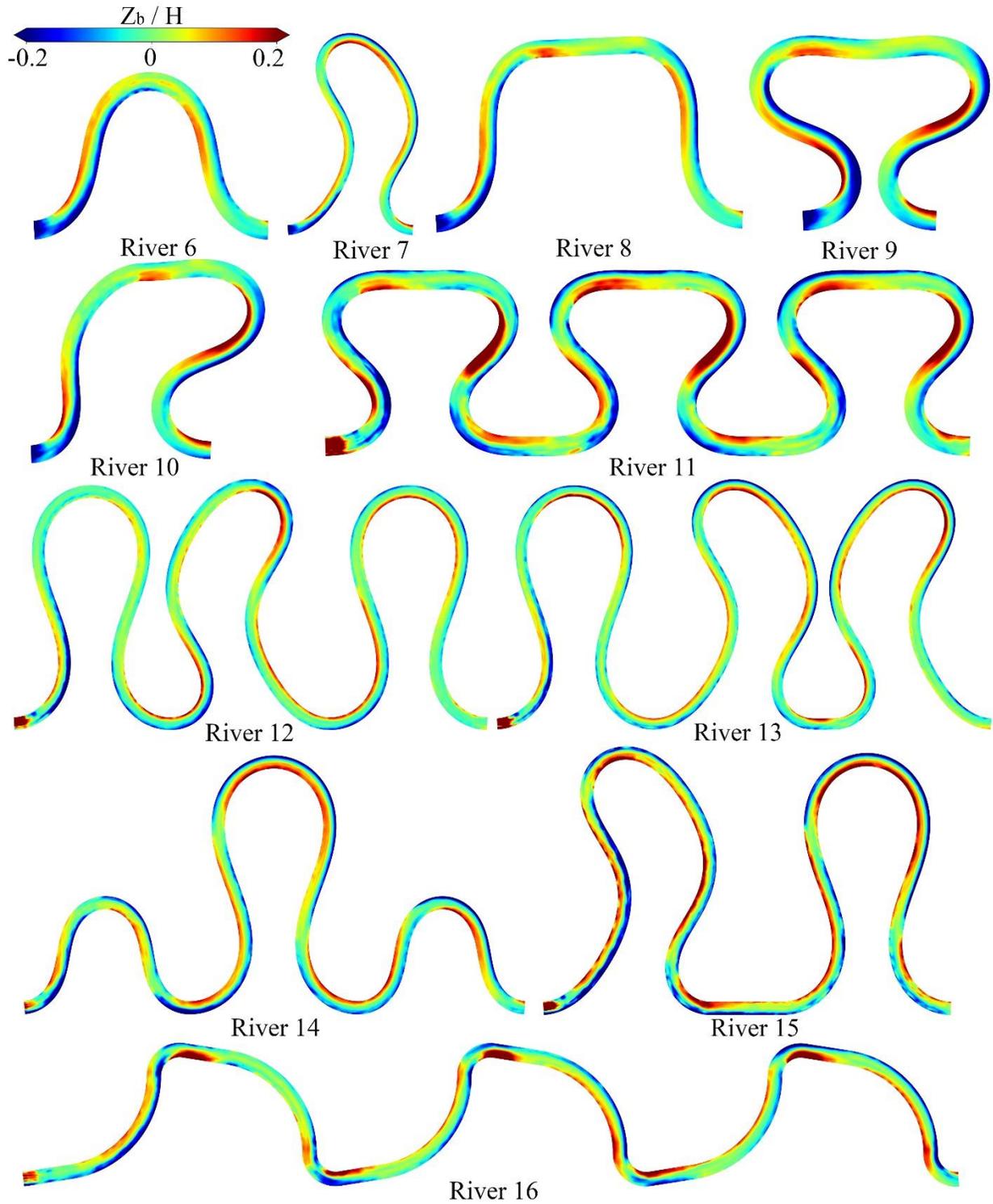

**Figure 15**: CNNAE predictions of the bed topography at equilibrium, non-dimensionalized by the river depth H.

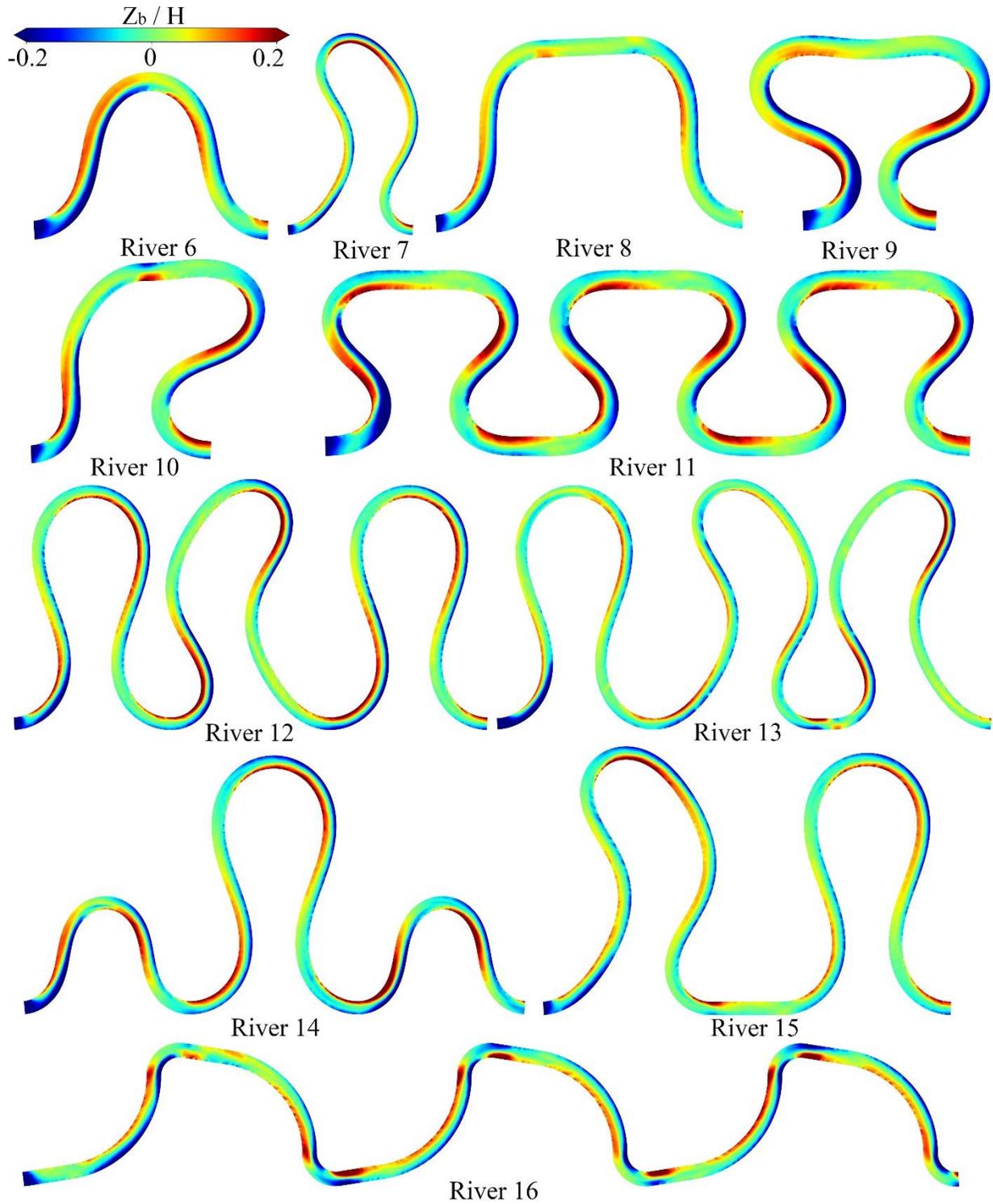

**Figure 16**: The coupled LES-morphodynamics model's computed bed topography at equilibrium, non-dimensionalized by the river depth H.

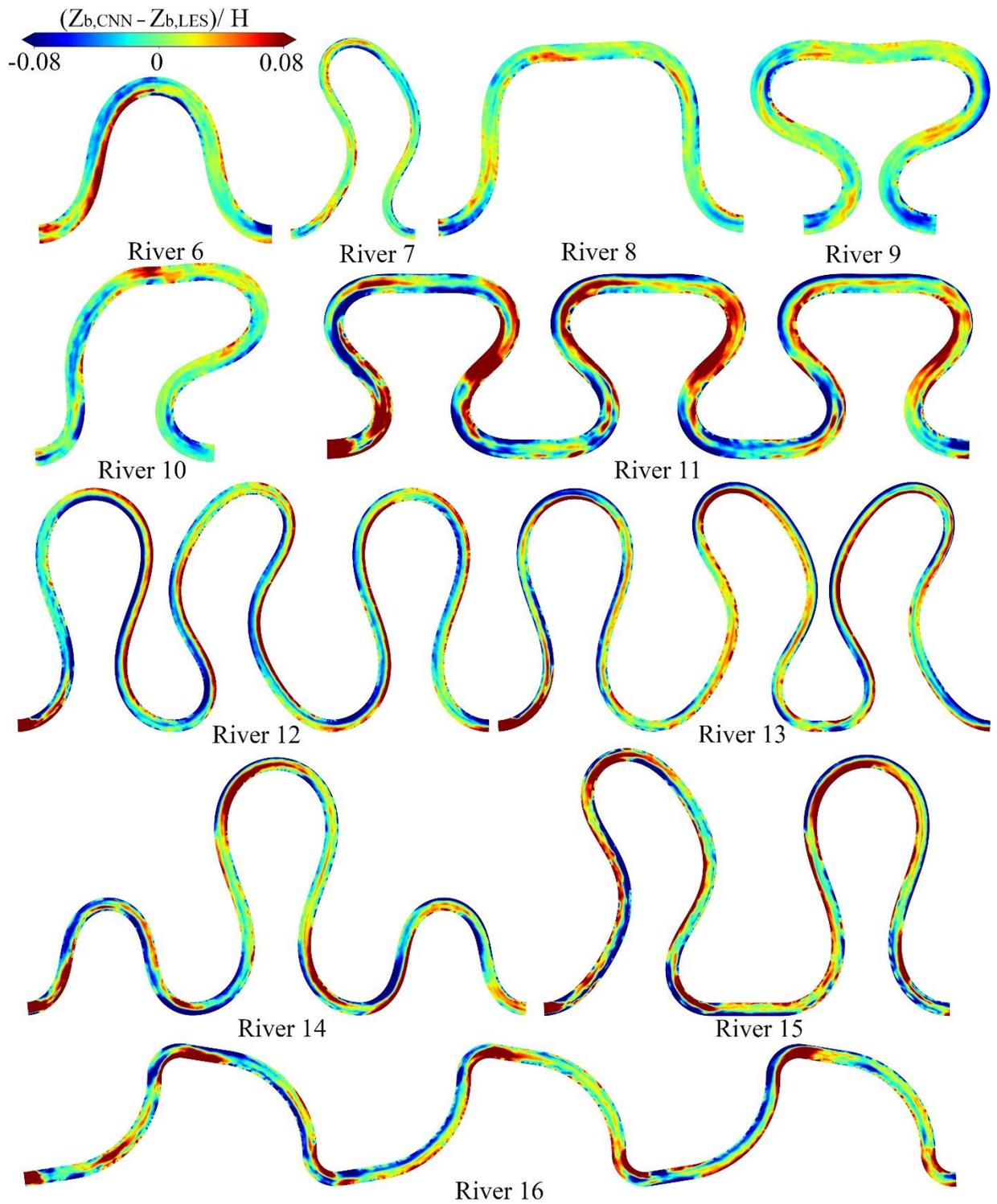

**Figure 17**: The prediction error of the equilibrium bed elevations, non-dimensionalized by the water depth H, between the coupled simulations and the CNNAE results.

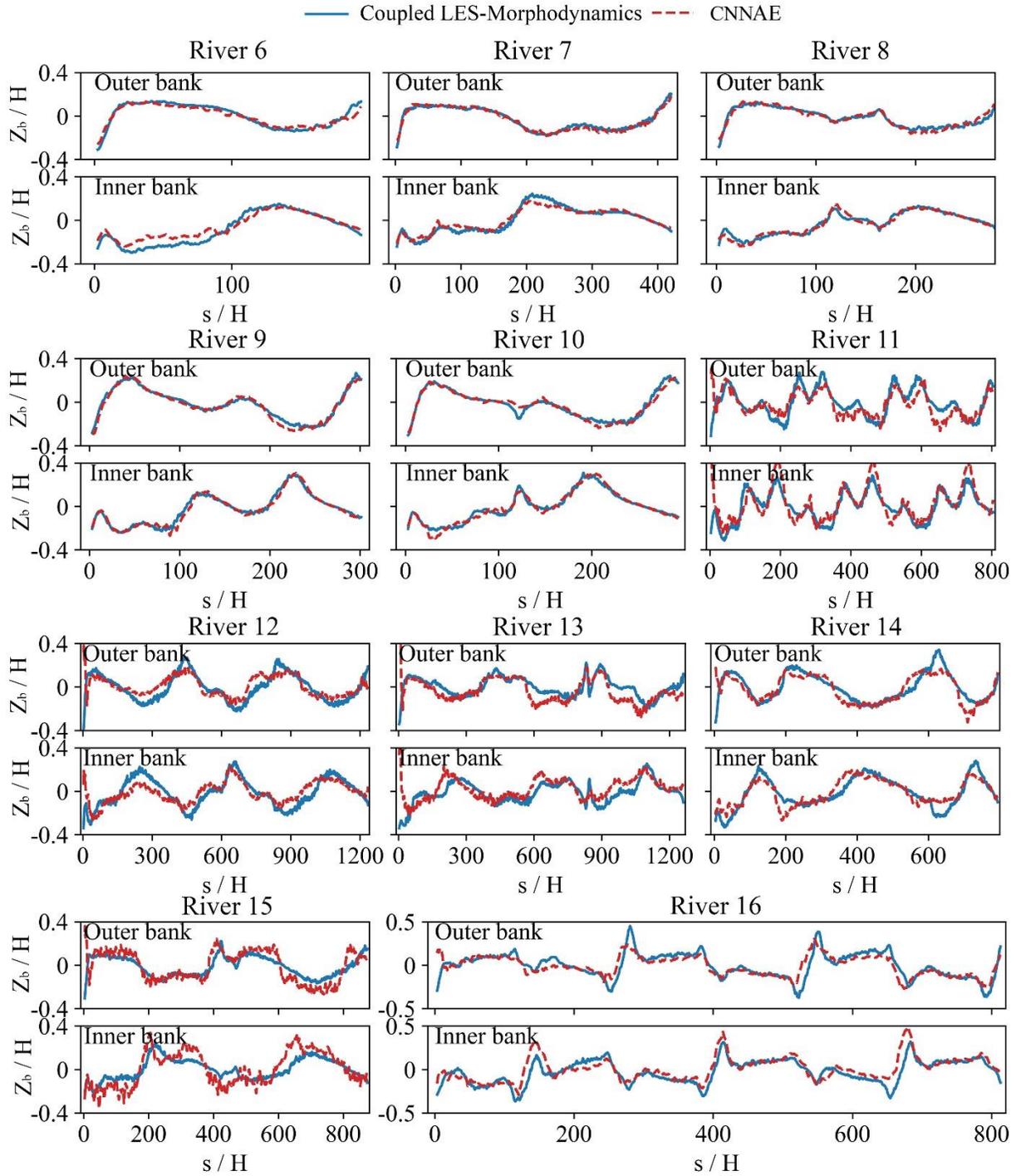

**Figure 18**: The predicted profiles of the time-averaged bed elevation normalized by water depth H. Profiles are extracted from streamwise lines 0.2W away from the outer and inner banks. "s" is the vector along the river. Solid blue and dashed red lines represent the CNNAE and the coupled LES-morphodynamics model results.

# 6. Conclusion

This paper proposes a novel CNNAE algorithm to predict the time-averaged shear stress distribution and equilibrium bed topography of mobile riverbed in large-scale meandering rivers under bankful flow conditions. We performed a series of coupled hydrodynamics and morphodynamics simulations in sixteen virtual meandering rivers with different geometry to generate training and validation data for the CNNAE algorithm. The CNNAE algorithm requires, as input vector, instantaneous snapshots of shear stress distribution and bed elevation change from the very beginning of the coupled hydrodynamics-morphodynamics simulation. The CNNAE also uses river's geometry parameters (e.g., streamwise and spanwise curvilinear coordinates, and local curvature) as the input, to reconstruct the time-averaged shear stress distribution and equilibrium bed elevation over the deformed riverbeds. The implemented CNNAE algorithm was trained using datasets from five 5 meandering rivers and validated against the high-fidelity CFM model's results in eleven large-scale meanders. The following conclusions can be drawn from this study:

(*i*) The trained CNNAE accurately reconstructs the high and low shear stress regions of the meanders. It also captures scour and deposition regions of the deformed riverbeds at equilibrium. Considering that it uses the instantaneous snapshots from the nearly flat riverbed, the performance deems impressive.

(*ii*) Successful prediction of the shear stress distributions over the deformed riverbeds using the input vectors that originated form near flat riverbeds marked the effectiveness of the river geometry parameters introduced as inputs to CNNAE.

(*iii*) Successful prediction of the bed topography showed that the implemented CNNAE can recognize patterns of instantaneous elevation change, $\Delta Z_b$, to reconstruct equilibrium bed topography, while the two have no similarity in their pattern.

(*iv*) Compared to the high-fidelity coupled simulations, the proposed AI approach is highly efficient as it can generate high-fidelity topography data with less than 2% of the computational cost (i.e., CPU hours using the same computing cluster) required by the former. This demonstrates the potential of the proposed algorithm to carry out studies on the morphodynamics of field-scale meandering rivers.